\documentclass[lettersize,journal]{IEEEtran}

\usepackage{graphicx}
\usepackage{epstopdf}
\usepackage{booktabs} % For formal tables

\usepackage[ruled,vlined]{algorithm2e}
\usepackage{algpseudocode}

\usepackage{bm}
\usepackage{amsmath}

\usepackage{amssymb}
\usepackage{multirow}
\usepackage{epsfig}
\usepackage{xspace}
\usepackage{url}
\usepackage{color}
\usepackage{booktabs}
\usepackage{multirow}
\usepackage{makecell}
\usepackage{indentfirst}
\usepackage{subfigure}
\usepackage{xcolor}
\usepackage{cite}
\usepackage{graphicx}
\usepackage{hyperref}
\usepackage{ragged2e}

\usepackage{enumitem}
\usepackage{flushend}

\newtheorem{myDef}{Definition}

\newcommand{\paratitle}[1]{\vspace{1.5ex}\noindent\textbf{#1}}
\newcommand{\ie}{\emph{i.e.,}\xspace}

\newcommand{\etc}{\emph{etc.}\xspace}
\newcommand{\name}{ST-CLIP\xspace}
\newcommand{\prompt}{SCAMP\xspace}
\newcommand{\ignore}[1]{}
 
\begin{document}

\title{Spatio-Temporal Data Enhanced Vision-Language Model for Traffic Scene Understanding}

\author{Jingtian~Ma,~
        Jingyuan~Wang*,~%~\IEEEmembership{Member,~IEEE,}
        Wayne~Xin~Zhao,~%~\IEEEmembership{Member,~IEEE,}
        Guoping~Liu,~
        Xiang~Wen

\IEEEcompsocitemizethanks{
%\IEEEcompsocthanksitem * Corresponding author.
\IEEEcompsocthanksitem J. Ma is with the School of Computer Science and Engineering, and the MOE Engineering Research Center of Advanced Computer Application Technology, Beihang University, Beijing, China. E-mail: majingtian@buaa.edu.cn \protect
\IEEEcompsocthanksitem J. Wang is with the School of Computer Science and Engineering, the School of Economics and Management, and the MIIT Key Laboratory of Data Intelligence and Management, Beihang University, Beijing, China. E-mail: jywang@buaa.edu.cn
\IEEEcompsocthanksitem W.X. Zhao is with Gaoling School of Artificial Intelligence, Renmin University of China, Beijing, China.\protect
\IEEEcompsocthanksitem G. Liu and X. Wen are with DiDi Global Inc., Beijing, China. \protect 
\IEEEcompsocthanksitem  J. Wang is the corresponding author.
}
}

\markboth{Journal of \LaTeX\ Class Files,~Vol.~14, No.~8, August~2021}%
{Shell \MakeLowercase{\textit{et al.}}: A Sample Article Using IEEEtran.cls for IEEE Journals}


\maketitle

\begin{abstract}
Nowadays, navigation and ride-sharing apps have collected numerous images with spatio-temporal data. 
A core technology for analyzing such images, associated with spatio-temporal information, is Traffic Scene Understanding (TSU), which aims to provide a comprehensive description of the traffic scene.
Unlike traditional spatio-temporal data analysis tasks, the dependence on both spatio-temporal and visual-textual data introduces distinct challenges to TSU task.
However, recent research often treats TSU as a common image understanding task, ignoring the spatio-temporal information and overlooking the interrelations between different aspects of the traffic scene. 
To address these issues, we propose a novel \emph{\underline{S}patio-\underline{T}emporal Enhanced Model based on \underline{CILP}} (\name) for TSU. 
Our model uses the classic vision-language model, CLIP, as the backbone, and designs a Spatio-temporal Context Aware Multi-aspect Prompt (SCAMP) learning method to incorporate spatio-temporal information into TSU. 
The prompt learning method consists of two components: A dynamic spatio-temporal context representation module that extracts representation vectors of spatio-temporal data for each traffic scene image, and a bi-level ST-aware multi-aspect prompt learning module that integrates the ST-context representation vectors into word embeddings of prompts for the CLIP model. 
The second module also extracts low-level visual features and image-wise high-level semantic features to exploit interactive relations among different aspects of traffic scenes.
To the best of our knowledge, this is the first attempt to integrate spatio-temporal information into vision-language models to facilitate TSU task.
Experiments on two real-world datasets demonstrate superior performance in the complex scene understanding scenarios with a few-shot learning strategy.
\end{abstract}

\begin{IEEEkeywords}
Traffic Scene Understanding, Spatio-Temporal Data, Prompt Learning
\end{IEEEkeywords}

\section{Introduction}\label{sec:introduction}

\IEEEPARstart{W}ith the advancement of onboard cameras and GPS-equipped devices, transportation service systems have collected numerous image data with spatio-temporal information, crucial for many intelligent transportation applications like traffic flow prediction~\cite{ji2023spatio, jiang2023pdformer, han2025bridging, ji2025seeing, liu2024full}, autonomous driving~\cite{chen2023driving, mao2023language, yurtsever2020survey}, and route recommendation~\cite{wang2019empowering, 9385890, dai2015personalized, guo2020force, wu2019learning}. 
\emph{Traffic scene understanding} (TSU), which aims to provide intuitive semantic information, usually in text format, for understanding the driving environment based on complex multimodal data~\cite{li2020traffic}, is a core technology to analyze the image data associated with spatio-temporal information.
Unlike traditional traffic-related tasks, TSU depends on both spatio-temporal context and visual-textual data to provide a comprehensive description of the traffic scene, which makes it more challenging than traditional transportation data processing tasks.
In this paper, we aim to find a solution that combines pre-trained visual-language models and spatio-temporal data to further optimize the TSU task.

In the literature, to solve the task of TSU, early research relies on deep learning-based visual methods, such as image/video classification~\cite{Narayanan2019DynamicTS, bilen2017action}, object detection~\cite{zhu2016traffic, guindel2018fast, can2021structured} and semantic segmentation~\cite{chen2020pixelwise, cordts2016cityscapes, behley2019semantickitti}, to identify key elements in traffic scenes. 
However, these methods focus exclusively on specific low-level features of traffic scene images, such as counting the number of vehicles and distinguishing between lanes and curbs, which fails to capture the high-level relationships between different aspects and provide comprehensive descriptions of the traffic scenes, such as the overall scene in the traffic images, the condition and accessibility of the road, and other relevant factors.
Additionally, the aforementioned models for the TSU task require a substantial amount of annotated data for training. The limited availability of descriptive text labels for traffic scene images further restricts the accuracy and comprehensiveness for addressing the TSU task.

\begin{figure}[t]
   \centering
   \includegraphics[width=0.98\columnwidth]{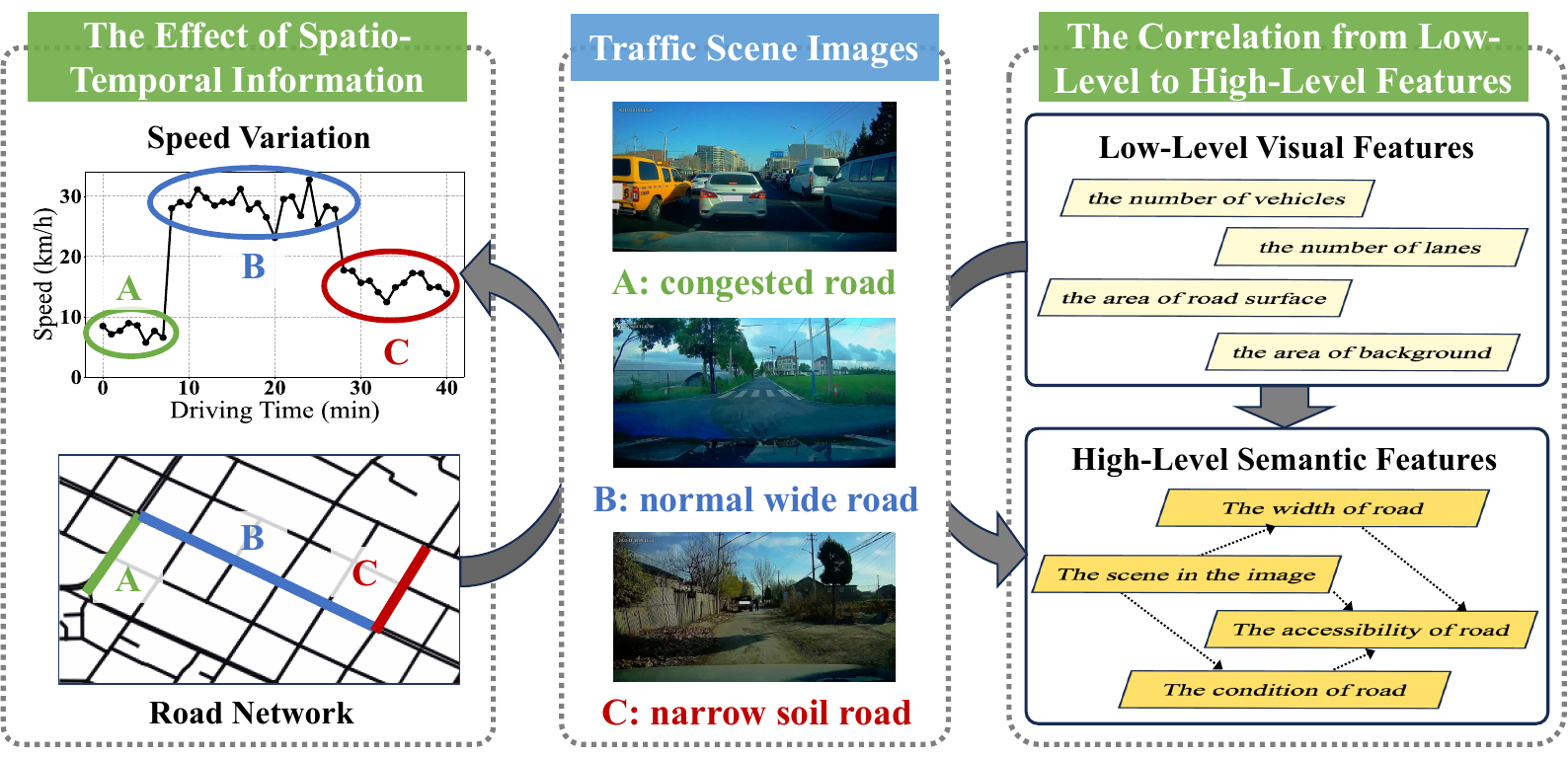}
   \caption{Two main issues that are overlooked with VLMs for the TSU task.}
   % \ie given the traffic scene image and its corresponding spatio-temporal context, providing the description of the scene.}
   \label{fig:drawbacks}
\end{figure}

In recent years, pre-trained large-scale Vision-Language Models (VLMs) have demonstrated impressive capabilities in highly challenging multimodal tasks, such as image captioning~\cite{yang2024exploring, xu2024pllava}, visual question-answering~\cite{wang2023fashionvqa, zhang2024can}, and image-text retrieval~\cite{huang2024cross, zhai2022lit}. 
Equipped with extensive parameters and trained on massive datasets, VLMs provide strong few-shot transferability and excellent scalability, enabling effective adaptation to new TSU tasks with minimal labeled data and thus avoiding the costly burden of large-scale annotations.
For different applications, VLMs are typically guided by a piece of text, \ie a prompt, to instruct the model on what task to perform. 
With a guided prompt, VLMs can be easily adapted to various downstream tasks, requiring only a small amount of annotated data to achieve good performance.
Therefore, VLMs are naturally suitable for TSU tasks. 
By inputting the traffic scene image with a prompt specifically designed for traffic scenes, a VLM can directly generate a comprehensive paragraph of text description for the image.

However, there are at least two major issues that have not been well addressed in directly using VLMs to solve the TSU task.
Firstly, existing VLMs largely overlook the importance of spatio-temporal information in traffic scene understanding. 
With the proliferation of GPS-enabled devices, spatio-temporal data such as vehicle trajectories and real-time traffic flows provide critical insights into traffic dynamics and human behaviors. For example, as shown in the left part of Fig.~\ref{fig:drawbacks}, average speeds vary significantly across congested roads, wide roads, and narrow soil roads, reflecting distinct traffic conditions. Additionally, changes in driving behavior, such as sudden acceleration and deceleration, can signal shifts in the traffic scenes, such as obstacles or road construction. Ignoring these spatio-temporal cues limits VLMs' ability to fully capture real-world traffic complexity.
Secondly, existing methods primarily focus on low-level visual features, such as vehicle counts or lane markings, which offer only a superficial understanding of traffic scenes. To achieve a comprehensive analysis, it is essential to explore high-level semantic features and their interrelationships. For instance, in the right part of Fig.~\ref{fig:drawbacks}, a narrow soil road (C) is more likely to pose challenges for navigation compared to a broad and smooth road (B). Here, both width and surface of a road serve as key factors that influence the road's accessibility. Analyzing these high-level features and their correlations is crucial for accurately interpreting traffic scenes, as isolated low-level features fail to capture the full context.

To address these issues, we propose a novel \emph{\underline{S}patio-\underline{T}emporal Data Enhanced \underline{CLIP}} model, denoted as \emph{\name}, for the TSU task. 
We leverage prompt learning to integrate spatio-temporal data with pre-trained vision-language models and design a bi-level multi-prompt attention mechanism to address the issue of insufficient high-level semantic modeling. 
Specifically, the \name model uses the classic vision-language model, Contrastive Language-Image Pre-training (CLIP) model, as the backbone. 
The inputs to \name include a traffic scene image and associated spatio-temporal data, such as traffic state and vehicle trajectories. 
We propose a Spatio-temporal Context Aware Multi-aspect Prompt (\prompt) method to learn appropriate prompts for the backbone VLM based on spatio-temporal additional data. 
In the \prompt method, we first design a Dynamic Spatio-temporal Context Representation module to extract the features of real-time traffic state, vehicle location trajectories, and correlations among locations in a trajectory as ST-context representation vectors of each traffic scene image. 
Then, we propose an automatic prompt learning method to construct learnable word embeddings for prompt words based on these ST-context representation vectors. 
Utilizing a bi-level multi-aspcet prompt attention mechanism to facilitate knowledge sharing, our approach generates multi-aspect prompts that integrate both patch-wise low-level visual features and image-wise high-level semantic features of traffic scene images. 
Finally, we adopt a cross entropy loss function to train the parameters of the \prompt method. 
Using a few-shot learning approach, our \prompt method can generate effective prompts for the backbone vision-language model.
Based on that, our \name model can fully leverage the pre-trained model's rich knowledge and integrate spatio-temporal context information for comprehensive multi-aspect traffic scene understanding.

The main contributions of this paper are as follows:
\begin{itemize}
    \item To the best of our knowledge, this is the first attempt to integrate spatio-temporal information into pre-trained multimodal models to facilitate the task of TSU. 
    \item We propose a dynamic ST-context representation method that captures comprehensive and dynamic spatio-temporal characteristics, enabling ST-context aware prompts for large multimodal models.
    \item We introduce a bi-level multi-aspect prompt attention mechanism that simultaneously models low-level cross-modal features and high-level cross-aspect correlations, enhancing complex traffic scene understanding.
    \item Extensive experiments on two real-world datasets demonstrate the effectiveness of our “pre-trained model + spatio-temporal data” framework, which can be extended to other traffic-related applications. 
\end{itemize}

\section{RELATED WORK}~\label{sec:related_works}
\vspace{-0.3cm}

Our work is related to the following research directions.

\vspace{-0.15cm}
\paratitle{Traffic Scene Understanding.} Traffic scene understanding refers to the task of automatically analyzing and comprehending traffic scenes, which involves various aspects such as object detection~\cite{zhu2016traffic, guindel2018fast, can2021structured}, semantic segmentation~\cite{cordts2016cityscapes, behley2019semantickitti} and road condition analysis~\cite{6082921, amirgaliyev2016road, ruotoistenmaki2007road, bhatt2017intelligent}.
Traditional approaches relied on GPS records and weather sensors~\cite{6082921, amirgaliyev2016road, ruotoistenmaki2007road, bhatt2017intelligent}, but due to their simplicity, these methods achieved limited performance.
With the rise of deep learning, more sophisticated spatio-temporal models were introduced. A seminal work, SAE~\cite{lv2014traffic}, pioneered deep learning applications in traffic flow prediction, highlighting the importance of jointly modeling temporal dependencies and spatial correlations. Building on this direction, graph-based methods~\cite{yanghygmap, zhang2024veccity, ji2020interpretable, wu2020learning, ji2022stden, wang2022traffic, jiang2023self} explicitly model road network topology with temporal dynamics, demonstrating the power of graph-based representations for traffic flow forecasting.
On the perception side, some studies focus on improving the accuracy of traffic object detection~\cite{zhu2016traffic, guindel2018fast, can2021structured}. Others aim to improve scene classification and segmentation accuracy~\cite{behley2019semantickitti, chen2020pixelwise, cordts2016cityscapes}.
Recent work has also linked TSU with image captioning~\cite{li2020traffic}, generating textual descriptions of traffic situations. However, most existing models focus primarily on static spatial cues and overlook the temporal variations of vehicle trajectories, which are essential for capturing the dynamic nature of traffic scenes.

\vspace{-0.15cm}
\paratitle{Vision-Language Models.} Recent years have witnessed significant progress in the field of multimodal learning~\cite{sun2019videobert, li2019visualbert, li2020oscar}.
Joint vision-language models have demonstrated impressive capabilities in highly challenging tasks such as image captioning~\cite{liu2021cptr, xu2018attngan}, visual question-answering~\cite{hu2018explainable, yi2018neural, zellers2019recognition, zhang2021explicit}, and image-text retrieval~\cite{long2022gradual, chen2024make}. 
Early work pioneered the use of natural language to guide visual understanding, laying the foundation for textual prompts in vision tasks~\cite{hu2016segmentation}.
A milestone in this area is CLIP~\cite{radford2021learning}, which learns to recognize paired image and text with a contrastive pre-training paradigm.
Specifically, given a batch of $N$ (image-text) pairs, the goal is to predict which of the $N$ × $N$ possible pairs are matched pairs (positive samples) and which are unmatched pairs (negative samples).
Equipped with large-scale (400M image-text pairs) web-crawled data for pre-training, CLIP transfers non-trivially to most tasks and is often competitive with a fully supervised baseline without the need for any dataset specific training.
The cross-modal knowledge in CLIP has inspired a large number of follow-up works, which typically introduce a few additional parameters for fine-tuning while keeping the CLIP model parameters fixed.
Some approaches adopt simple adapters to learn new features and achieve effective results~\cite{gao2021clip, zhang2021tip}.
While other methods adopt continuous prompt learning method to invoke the potential of CLIP, avoiding the manual design of prompts~\cite{zhou2022learning, zhou2022conditional}. 
However, for the TSU task, these models typically overlook the interrelationships between various aspects of the traffic scene. Additionally, they fail to integrate essential traffic domain knowledge, such as road properties and trajectory features, which are crucial for accurate scene understanding.

\vspace{-0.2cm}
\paratitle{Prompt Learning.} Prompt learning is a relatively new paradigm in natural language processing (NLP), which has received growing attention in recent years~\cite{cheng2025poi, shin2020autoprompt, khashabi2021prompt, yu2025bigcity, liu2021gpt, liu2022p}. 
It involves training a model to generate natural language text conditioned on a given prompt, which can be a short phrase or sentence. 
The representative work is GPT-4, which achieves state-of-the-art performance on a wide range of language tasks by training a large-scale transformer model to generate text from a given prompt. 
In the vision-language modeling, CoOp~\cite{zhou2022learning} introduced learnable prompts for CLIP, improving adaptability to downstream tasks. Its extension, CoCoOp~\cite{zhou2022conditional}, employed conditional prompts that dynamically adjust to each image, enhancing generalization. CLIP-Adapter~\cite{gao2021clip} achieved parameter-efficient tuning by inserting lightweight residual adapters, while Tip-Adapter and Tip-Adapter-F~\cite{zhang2021tip} further boosted few-shot adaptation by caching visual features and fusing them with CLIP predictions.
Our proposed model inherits the spirit of prompt learning but extends it in two key ways: introducing spatio-temporal priors to capture the dynamics of road networks, and modeling multi-aspect prompts jointly. This design is specifically tailored to the requirements of TSU task, going beyond generic prompt-tuning methods.
\vspace{-0.3cm}
\section{Preliminaries}~\label{sec:preliminaries}

\vspace{-1.05cm}
\subsection{Problem Definition}

The input of a traditional TSU model is a street scene image and the output is a descriptive text for the image. 
Compared with general image understanding applications, a characteristic of TSU is that the scene images are accompanied by spatio-temporal information.
Since in-vehicle cameras that capture the scene images are often equipped with GPS terminals, the precise location of vehicles and time of photo taken could be recorded with the scene images. 
In this way, we can use the spatio-temporal context as prior knowledge to achieve better traffic scene understanding. Therefore, we provide the following definition for the TSU problem.

\begin{figure}[t]
   \centering
   \includegraphics[width=0.9\columnwidth]{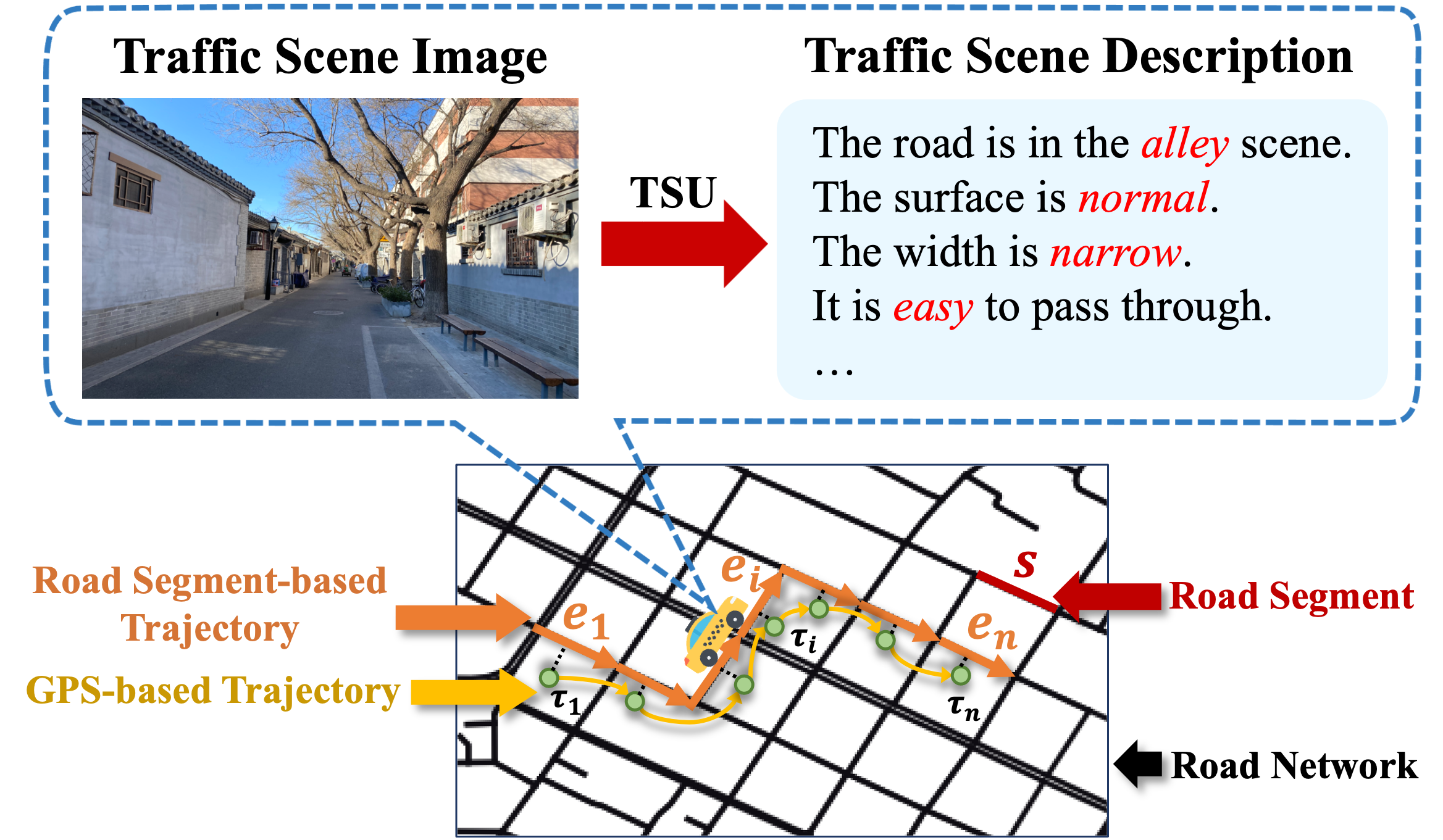}
   \caption{Illustration of the TSU task in this work, \ie given a traffic scene image and its corresponding spatio-temporal context, providing the description of the scene.}
   \label{illustration}
\end{figure}

\begin{myDef}[Traffic Scene Understanding]~\label{def:TSU}
Denoting $\mathcal{I}$ as a traffic scene image, and $\mathcal{R}$ as the spatio-temporal context information related to the image, a TSU model is defined as
\begin{equation}\label{}
    \mathcal{D} = \mathrm{TSU}(\mathcal{I}, \mathcal{R}),
\end{equation}
\end{myDef}
where $\mathcal{D}$ is the text describing the traffic scene image.
Fig.~\ref{illustration} is an illustration of TSU task, where the inputs are a traffic scene image and the spatio-temporal information of the vehicle when capturing the image, and the output of the model is a textual description which depicts the traffic scene in the image. 

\vspace{-0.45cm}
\subsection{CLIP-based Scene Understanding}

The \name model is based on the classic Contrastive Language-Image Pre-training model, namely CLIP~\cite{radford2021learning}, which is a vision-language model pre-trained for general image understanding. 
The CLIP model consists of an image encoder and a text encoder, which are respectively utilized to convert images and descriptive texts into feature vectors. 
% Given an image and its descriptive text
Furthermore, the CLIP model aligns the visual and textual feature vectors using a contrastive learning strategy, which enables the model to bridge the gap between visual and textual information.

\paratitle{Image and Text Encoders.} Given an image $\mathcal{I}$, the image encoder of CLIP converts it into a feature vector 
$\bm{i}\in \mathbb{R}^{D}$, \ie
\begin{equation}\label{eq:img_encoder}
  \bm{i} = \mathrm{Encoder_{img}}(\mathcal{I}),
\end{equation}
where $\mathrm{{Encoder}_{img}}(\cdot)$ is typically implemented using a CNN-based architecture such as ResNet-50 \cite{he2016deep} or a Vision Transformer (ViT) \cite{dosovitskiy2020image}. 
Meanwhile, given the descriptive text of an image, the CLIP model first uses an embedding layer to embed each word of the text as a vector and then uses a text encoder to convert the word embedding sequence of the image description, denoted as $\bm{T}$, into a feature vector $\bm{t}\in \mathbb{R}^{D}$ as
\begin{equation}\label{eq:txt_encoder}
  \bm{t} = \mathrm{Encoder_{txt}}(\bm{T}),
\end{equation}
where $\mathrm{Encoder_{txt}}(\cdot)$ is implemented by a Transformer \cite{vaswani2017attention}.

\paratitle{Contrastive Learning Pre-training.} In the CLIP model, the image and text encoders are pre-trained using a contrastive learning strategy with image-text pairs dataset. 
The training data is a set of images and their descriptive texts. 
CLIP considers these matched image-text pairs as positive samples while randomly combines images and texts as unmatched negative samples. 
The objective of the contrastive learning strategy is to maximize the cosine similarity of image and text feature vectors for positive samples while minimize the cosine similarity for unmatched negative samples. 
Specifically, given $X$ image-text pairs, denoting $(\bm{i}_x, \bm{t}_x)$ as feature vectors of a matched image-text pair, and $(\bm{i}_x, \bm{t}_y)$ where $x \neq y$ as an unmatched negative sample, the loss function of the contrastive learning strategy in CLIP is defined as
\begin{equation}\label{}
  L_{c} = \sum_{x=1}^{X} -\log \frac{\exp(\cos(\bm{i}_x,\bm{t}_x)/\mu)}{\sum_{y\neq x}\exp(\cos(\bm{i}_x,\bm{t}_y)/\mu)},
\end{equation}
where $\exp(\cdot)$ represents the exponential function, $\cos(\cdot,\cdot)$ denotes the cosine similarity, and $\mu$ is the temperature parameter,
which controls the concentration of probability distribution. 
To acquire a comprehensive range of visual concepts and enhance the transferability of the acquired knowledge to diverse applications, the CLIP team collected an extensive training dataset comprising 400 million matched image-text pairs~\cite{radford2021learning}.

\paratitle{CLIP-based Scene Understanding.} After pre-trained on huge training dataset with a constrastive learning strategy, CLIP inherits strong zero-shot transfer capability for diverse downstream applications. 
Based on CLIP, we construct a basic image scene understanding framework. 
Specifically, we assume there are $K$ classes of scenes in the images. For each class of scene, we define a class-specific word, denoted as  $([\mathrm{CLASS}]_1, \ldots, [\mathrm{CLASS}]_k, \ldots, [\mathrm{CLASS}]_K)$. For the $k$-th class-specific word, we construct a prompt as
\begin{equation}
    \mathrm{PMPT}_k=[\mathrm{W}]_{1}\dots[\mathrm{W}]_{m}\ldots[\mathrm{W}]_{M}[\mathrm{CLASS}]_k,
\end{equation}
where $[\mathrm{W}]_m$ is the $m$-th word of the prompt.
The CLIP model first converts the prompt into a embedding vector sequence:
\begin{equation}
    \bm{T}_k=\left(\bm{v}_1,\dots,\bm{v}_m,\ldots,\bm{v}_M,\bm{c}_k\right),
\end{equation}
where $\bm{v}_{m}$ is the embedding vector for the word $[\mathrm{W}]_{m}$, and $\bm{c}_k$ is the embedding vector for the class-specific word $[\mathrm{CLASS}]_k$. Then, the CLIP model converts $\bm{T}_k$ as a feature vector, denoted as $\bm{t}_k$, using the text encoder in Eq.~\eqref{eq:txt_encoder}.

Given the textual feature vectors for corresponding $K$ prompts, denoted as $(\bm{t}_1, \ldots, \bm{t}_k, \ldots, \bm{t}_K)$, and the visual feature vector $\bm{i}$ of the image to be understood, the index of the matched class-specific word is calculated as
\begin{equation}\label{eq:inference}
     k* = \mathop{\arg\max}\limits_{k} \frac{\mathrm{exp}(\mathrm{cos}(\bm{i}, \bm{t}_k)/\mu)}{\sum_{j=1}^K\mathrm{exp}(\mathrm{cos}(\bm{i}, \bm{t}_j)/\mu)},
\end{equation}
The corresponding prompt, such as \textit{a photo of a road which is [CLASS]$_{k*}$}, is the scene understanding output.

\vspace{-0.2cm}
\section{Spatio-temporal Context Aware Multi-aspect Prompt Learning}~\label{sec:prompt}

\begin{figure*}
    \centering
    \includegraphics[width=0.85\linewidth]{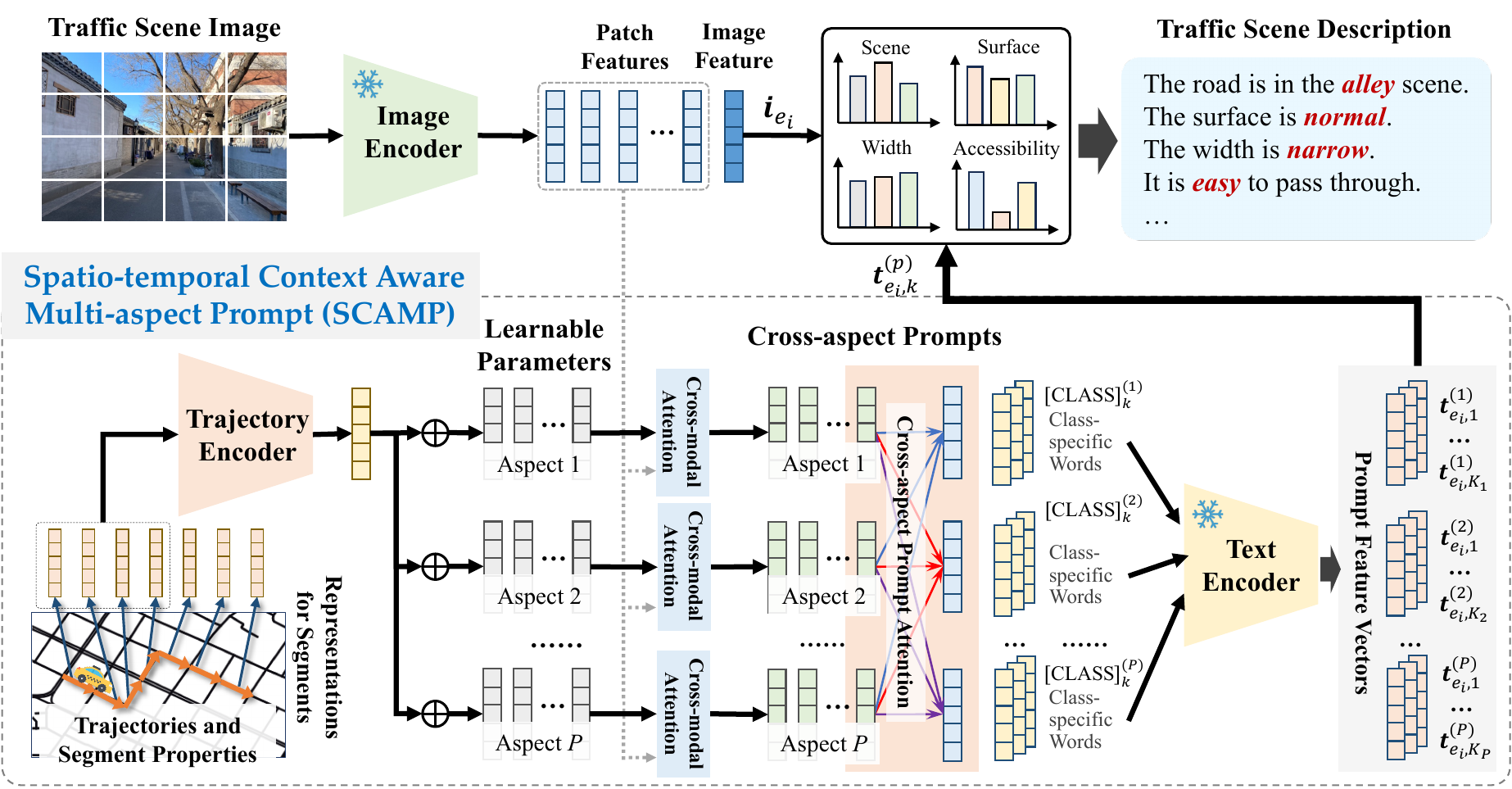}
    \caption{The overall framework of ST-CILP with Spatio-temporal Context Aware Multi-aspect Prompt (SCAMP) learning method in detail.}
    \label{fig:framework}
\end{figure*}

\vspace{-0.4cm}
In this section, we propose a Spatio-Temporal data enhanced CLIP~(ST-CLIP) model with Spatio-temporal Context Aware Multi-aspect Prompt~(SCAMP) learning method in this section. The overall framework is shown in Fig.~\ref{fig:framework}. 
We start with the representation learning of dynamic spatio-temporal context, then present the construction of learnable ST-aware multi-aspect prompts and the bi-level attention mechanism, and finally introduce how to train the entire model and generate the traffic scene descriptions. 

\vspace{-0.4cm}
\subsection{Dynamic Spatio-temporal Context Representation}\label{sec:stRL}

\subsubsection{Road Segment-based Trajectory}

In the TSU task, the spatio-temporal information accompanying the image consists of the vehicle's trajectories at the time the image was captured. To facilitate data processing, we match the precise locations of the trajectories to the road network and use road segment-based trajectories as auxiliary information to generate dynamic spatio-temporal context aware prompts.

We first define the road segment and road network.
\begin{myDef}[Road Segment and Road Network]\label{def:road_segment}
    A road segment $s_i$ is a uniform section of road, which is the basic location unit in city map. For each segment, there is a set of properties, including road segment ID, function class, lane number, speed class, road length and \etc %We denote the properties of the segment $s_i$ as $\mathcal{V}_i$.
    A road network is characterized as a directed graph $\mathcal{G}=\langle \mathcal{S},\bm{A} \rangle$, where $\mathcal{S}$ is a vertex set of road segments and $|\mathcal{S}| = N_s$. $\bm{A} \in \mathbb{R}^{N_s \times N_s}$ is the adjacency matrix of the graph. If two segments are connected, the corresponding entry of $\bm{A}$ is {\color{blue}1}, otherwise is 0. 
\end{myDef}

A raw GPS-based trajectory generated by a vehicle is a sequence of coordinates with timestamps. We uses the Fast Map Matching (FMM) algorithm~\cite{yang2018fast} to convert raw trajectories into road segment-based trajectories.

\begin{myDef}[Road Segment-based Trajectory]
A road segment-based trajectory is a sequence of connected road segments with timestamps, denoted as $\bm{tr}=(e_1, \ldots, e_i, \ldots, e_I)$, where $e_i =  <s_{\tau_i}, \tau_i>$ is a segment sample, $s_{\tau_i}$ is the $i$-th road segment passed by the vehicle, and $\tau_i$ is the timestamp when the vehicle passed the $i$-th road segment.
\end{myDef}

Fig.~\ref{illustration} provides an example where each raw point coordinate corresponds to a road segment after mapping a GPS-based trajectory onto the road network, forming a road segment-based trajectory. Moreover, due to variations in passing time $\tau_i$, different trajectories exhibit distinct attributes when traversing the same road segment, indicating dynamic traffic scenes.

\subsubsection{Time-varying Representations for Segments}

In road segment-based trajectories, a road segment has two types of properties, \ie static properties, which are invariable with timestamp $\tau_i$, and time-varying properties, which change with $\tau_i$. Our model encode these two types of properties as a time-varying representation for each road segment in a trajectory.

\vspace{-0.15cm}
\paratitle{Static Properties.} For each segment, we consider six kinds of properties in its static properties encoding, namely ID, Function Class (FC), Lane Number (LN), Speed Class (SC), Road Length (RL), and Out Degree (OD). For discrete properties, including ID, FC, LN, SC, and OD, we use one-hot encoding with a learnable lookup matrix to generate their embedding vectors. For the continuous property, RL, we discretize its entire value range into several consecutive bins and utilize the bin number for coding. Then, the embedding vector of static properties for the segment $s$ is calculated as
\begin{equation}
    \bm{h}_{s}^{(s)} = \bm{h}_{ID} \Vert \bm{h}_{FC} \Vert \bm{h}_{LN} \Vert \bm{h}_{SC} \Vert \bm{h}_{RL} \Vert \bm{h}_{OD},
    \label{eq:hs}
\end{equation}
where ``$\Vert$'' denotes the vector concatenation operation and $\bm{h}_{(\cdot)}$ represents the embedding vector for corresponding properties.

\vspace{-0.15cm}
\paratitle{Time-varying Properties.} When we match a segment into a trajectory, the segment becomes an instance of the sample point for the trajectory. In this condition, some properties of the segment become variable with the sampling timestamp $\tau_i$, such as traffic conditions. In our model, we calculate two types of time-varying properties to capture the variable features for each segment, namely Trajectory Count (TC) and Medium Speed (MS). These time-varying properties are calculated from trajectories passing through a segment at the timestamp $\tau_i$. 
For the segment sample $e_i$ of a trajectory, we calculate the embedding vector of the dynamic properties as
\begin{equation}
    \bm{h}_{e_i}^{(d)} = \bm{h}_{TC} \Vert \bm{h}_{MS}.
    \label{eq:hd}
\end{equation}
Then, we utilize a feed-forward layer to transform the embedding vectors of static and time-varying properties into a comprehensive feature vector for the segment sample $e_i$ as 

\begin{equation}\label{eq:initial_feature}
    \bm{h}^{(0)}_{e_i} = \mathrm{FFN}\left(\bm{h}_{s}^{(s)}\Vert\bm{h}_{e_i}^{(d)}\right),
\end{equation}
where $\mathrm{FFN}(\cdot)$ is a fully connected network, and $\bm{h}^{(0)}_{e_i} \in \mathbb{R}^D$. 

\subsubsection{Trajectory-level ST-Context Representation}

We use a transformer-based model to convert the time-varying representation of segment samples in a trajectory into a trajectory-level spatio-temporal context representational vector. Since the traffic scene of a road segment is more related to its surrounding area rather than distant roads, we utilize a sliding window to divide a complete trajectory into tracklets and construct tracklet representations for traffic scene images.

\begin{myDef}[Tracklet Representation]
Given a trajectory $\bm{tr}=(e_1, \ldots, e_i, \ldots, e_I)$, its corresponding embedding vector sequence is denoted as $(\bm{h}_{e_1}^{(0)}, \dots, \bm{h}_{e_i}^{(0)}, \dots, \bm{h}_{e_I}^{(0)})$. If there is a traffic scene image taken at the sample $e_i$, we define a tracklet representation matrix $\bm{H}^{(0)}\in\mathbb{R}^{(2N_w+1)\times D}$ for the image as
\begin{equation}\label{eq:tracklet_representation}
    \bm{H}^{(0)}=\left(\bm{h}^{(0)}_{e_{i-N_w}}, \dots, \bm{h}_{e_{i}}^{(0)}, \dots, \bm{h}^{(0)}_{e_{i+N_w}}\right)^\top,
\end{equation}
where $N_w$ represents the preset window size. In this way, we transform the variable-length trajectory sequences into fixed-length tracklets, enabling us to better focus on the information of road segments near the captured images.
\end{myDef}

We utilize a standard Transformer encoder~\cite{vaswani2017attention} with $L$ blocks to encode the tracklet representation as
\begin{equation}
    \bm{H}^{(L)}=\mathrm{Transformer}(\bm{H}^{(0)}).
    \label{eq:trajectory_encoder}
\end{equation}
% In the last layer, the output is denoted as $\bm{H}^{(L)}$. 
The row vector of $\bm{H}^{(L)}\in\mathbb{R}^{(2N_w+1)\times D}$ corresponding to trajectory sample $e_i$ is denoted as $\bm{r}_{e_i} \in \mathbb{R}^{D}$, \ie
\begin{equation}~\label{eq:trajectory_feature}
    \bm{r}_{e_i}=\left(\bm{H}^{(L)}[N_w, :]\right)^\top.
\end{equation}
We use $\bm{r}_{e_i}$ as the final {\em Spatio-temporal Context Representation} of the traffic scene image taken at the sample $e_i$.

\vspace{-0.4cm}
\subsection{Learnable ST-aware Multi-aspect Prompts}~\label{sec:STMP}
\vspace{-0.4cm}

A comprehensive understanding of traffic scenes usually involves multiple aspects, such as road surface, accessibility and so on. 
Using individual prompts for each aspect may neglect the interdependencies between different aspects, leading to conflicts in TSU task. 
To solve this problem, we first carefully select distinct aspects to depict the characteristics of the traffic scene, which includes the environment of the road (scene), the condition of the road surface (surface), the width of the road (width), and the difficulty of passing through the road (accessibility). For each aspect, we give several class-specific words shown in Table~\ref{tab:aspects}. Then, we propose adaptive multi-aspect prompts with ST-context as inputs.

\begin{table}
  \small
  \centering  \caption{Aspects and class-specific words for TSU.}\label{tab:aspects}
  \begin{tabular}{c|c}
  \toprule
  {\bf Aspects} & {\bf Class-specific Words} \\\midrule
  Scene &  field, vehicles, alley, stall, unknown\\
  Surface & normal, broken, soil, unknown \\
  Width & normal, narrow, extremely narrow, unknown \\
  Accessibility & easy, hard, extremely hard \\
  \bottomrule
\end{tabular}
\end{table}

\subsubsection{Learnable Prompt Based on ST-context}

For each aspect $p \in \{1,\dots,P\}$ of the traffic scene, the prompt in CLIP without the class-specific word is defined as
\begin{equation}
    \mathrm{PMPT}^{(p)}=[\mathrm{W}]_1^{(p)}\dots[\mathrm{W}]_m^{(p)}\ldots[\mathrm{W}]_M^{(p)}.
\end{equation}
with the corresponding prompt embedding matrix
\begin{equation}~\label{eq:prompt_embed_matrix}
    \bm{\mathcal{V}}^{(p)}=\left(\bm{v}_1^{(p)},\dots,\bm{v}_m^{(p)},\ldots,\bm{v}_M^{(p)}\right).
\end{equation}
In standard CLIP, $\bm{\mathcal{V}}^{(p)} \in \mathbb{R}^{M \times D}$ is determined by hand-crafted prompt tokens and remains static for all images.

Given the complexity and instability of hand-crafted prompt engineering, the \name model introduces dynamic prompts that combine ST-context representations with learnable parameters. Specifically, for each aspect $p$, we initialize a learnable embedding matrix $\bm{\mathcal{W}}^{(p)} \in \mathbb{R}^{M \times D}$:
\begin{equation}
    \bm{\mathcal{W}}^{(p)} = \left(\bm{w}_1^{(p)},\dots,\bm{w}_m^{(p)},\dots,\bm{w}_M^{(p)}\right),
\end{equation}
where each $\bm{w}_m^{(p)} \in \mathbb{R}^{D}$ is trainable. Given an image at trajectory sample $e_i$ with ST-context $\bm{r}_{e_i}$ (see Eq.~\eqref{eq:trajectory_feature}), the adaptive prompt embedding is defined as
\begin{equation}\label{eq:context2prompt}
  \tilde{\bm{v}}_{m,e_i}^{(p)} = \bm{w}_m^{(p)} + \bm{r}_{e_i}.
\end{equation}
The final adaptive prompt embedding matrix becomes
\begin{equation}~\label{eq:learnable_prompt}
    \tilde{\bm{\mathcal{V}}}_{e_i}^{(p)}=\left(\tilde{\bm{v}}_{1,e_i}^{(p)},\dots,\tilde{\bm{v}}_{m,e_i}^{(p)},\ldots,\tilde{\bm{v}}_{M,e_i}^{(p)}\right).
\end{equation}

Unlike the fixed $\bm{\mathcal{V}}^{(p)}$ in standard CLIP, the adaptive prompts $\tilde{\bm{\mathcal{V}}}_{e_i}^{(p)}$ vary with the dynamic ST-context $\bm{r}_{e_i}$. The learnable parameters $\bm{w}_m^{(p)}$ provide sufficient capacity to adapt across diverse traffic scenes, enabling \name to construct prompts that are both context-aware and flexible.

\subsubsection{Bi-level Multi-aspect Prompt Attention}

Traffic scene images contain both low-level visual cues (e.g., vehicles, lanes) and high-level semantics (e.g., environment, road width, accessibility). A single image feature cannot fully capture both levels simultaneously. Since local features are often patch-dependent while high-level semantics rely on global context, we design a bi-level multi-aspect prompt attention mechanism consisting of patch-wise cross-modal attention and image-wise cross-aspect attention.

\paratitle{Patch-wise Cross-modal Attention.}
Traffic scene images exhibit diverse local semantics (e.g., road vs. sky regions), which provides insights into boundary and object detection. To leverage such information, we employ ViT as the visual encoder and adjust Eq.~(\ref{eq:img_encoder}) as
\begin{equation}
    \bm{F}_p, \bm{i} = \mathrm{Encoder_{img}}(\mathcal{I}),
    \label{eq:vit}
\end{equation}
where $\bm{F}_p \in \mathbb{R}^{N_p \times D}$ denotes patch features with $N_p$ patches.

Considering that descriptions of different aspects may correspond to distinct patches, we adopt the patch-wise cross-modal attention mechanism. 
Specifically, for aspect $p$, the adaptive prompt embedding $\tilde{\bm{\mathcal{V}}}_{e_i}^{(p)}$ serves as the query, while $\bm{F}_p$ is used as key and value:
\begin{equation}
\bm{Q}^{(p)}=\tilde{\bm{\mathcal{V}}}_{e_i}^{(p)}\bm{W}_q^{(p)}, 
\bm{K}^{(p)}=\bm{F}_p\bm{W}_k^{(p)}, 
\bm{V}^{(p)}=\bm{F}_p\bm{W}_v^{(p)},
\end{equation}
where $\bm{W}_q^{(p)}$, $\bm{W}_k^{(p)}$, and $\bm{W}_v^{(p)}$ are learnable parameters for aspect $p$. Then we use the multi-head attention (MHA) layer~\cite{vaswani2017attention} to share features of prompt and visual features as
\begin{equation}
    \hat{\bm{\mathcal{V}}}_{e_i}^{(p)} = \mathrm{MHA}(\bm{Q}^{(p)}, \bm{K}^{(p)}, \bm{V}^{(p)}).
\end{equation}
In this way, the prompt feature matrix $\hat{\bm{\mathcal{V}}}_{e_i}^{(p)}\in\mathbb{R}^{M\times D}$ adaptively correlates the contents of different patches in the traffic scene images, which captures the low-level visual features.

\paratitle{Image-wise Cross-aspect Attention.}
Different aspects in a traffic scene are often correlated (e.g., wide roads generally imply higher accessibility and broken roads are often difficult to pass through). To exploit such dependencies, we compute attention between prompt embeddings of aspects $p$ and $q$:
\begin{equation}~\label{eq:attetions}
    \bm{\mathrm{ATT}}_{p,q}=\mathrm{softmax}\left(\frac{\hat{\bm{\mathcal{V}}}^{(p)}_{e_i}\bm{W}_A^{(pq)}\hat{\bm{\mathcal{V}}}^{{(q)}^\top}_{e_i}}{\sqrt{D}}\right),
\end{equation}
where $\bm{W}_A^{(pq)} \in \mathbb{R}^{D\times D}$ is a learnable parameter matrix between the $p$-th and $q$-th aspects. $\bm{\mathrm{ATT}}_{p,q} \in \mathbb{R}^{M\times M}$ is an attention matrix. 
The refined prompt embedding is obtained by aggregating across all aspects:
\begin{equation}~\label{eq:attention_prompt}
    {\bm{\mathcal{V}}}^{(p)}_{e_i}=\sum_{q=1}^P\bm{\mathrm{ATT}}_{p,q}\hat{\bm{\mathcal{V}}}^{(q)}_{e_i}.
\end{equation}
In this way, the prompt feature matrix $\bm{\mathcal{V}}^{(p)}_{e_i} \in \mathbb{R}^{M \times D}$ integrates knowledge from correlated aspects, therefore comprehensively capturing high-level semantics.

For the $k$-th class of aspect $p$, the refined textual input is formed by concatenating $\bm{\mathcal{V}}^{(p)}_{e_i}$ with the embedding of the class-specific token $[\mathrm{CLASS}]_k^{(p)}$:
\begin{equation}\label{eq:concat_text}
    \bm{T}^{(p)}_{e_i,k}=\left({\bm{\mathcal{V}}}^{(p)}_{e_i},\bm{c}^{(p)}_k\right),
\end{equation}
where $\bm{c}^{(p)}_k\in\mathbb{R}^D$ is the word embedding of $[\mathrm{CLASS}]_k^{(p)}$. 
Feeding $\bm{T}^{(p)}_{e_i,k}$ into the text encoder in Eq.~\eqref{eq:txt_encoder} produces the final ST-aware multi-aspect textual feature $\bm{t}^{(p)}_{e_i,k} \in \mathbb{R}^D$:
\begin{equation}\label{eq:text_feature}
  \bm{t}^{(p)}_{e_i,k} = \mathrm{Encoder_{txt}}\left(\bm{T}^{(p)}_{e_i,k}\right).
\end{equation}

\vspace{-0.5cm}
\subsection{Model Training and Description Generation}~\label{sec:model_training}
\vspace{-0.5cm}

\paratitle{Model Training.} The \name consists of a CLIP base model and a SCAMP extended module. 
To leverage the extensive knowledge of CLIP, we freeze the parameters of the base model and fine-tune the parameters of the SCAMP module.

Given a traffic scene image $\mathcal{I}$ taken at the trajectory sample $e_i$, whose visual feature generated by the image encoder is denoted as $\bm{i}_{e_i}$ (see Eq.~(\ref{eq:vit})), the SCAMP module and text encoder construct a text feature vector $\bm{t}^{(p)}_{e_i,k}$ (see Eq.~(\ref{eq:text_feature})) for the image. Then, the probability of the image corresponds to the $k$-th class in the $p$-th aspect is calculated as
\begin{equation}\label{eq:pred}
    \hat{y}_{\mathcal{I},k}^{(p)} = \frac{\exp\left(\mathrm{cos}(\bm{i}_{e_i}, \bm{t}_{e_i,k}^{(p)})/\mu\right)}{\sum_{k'=1}^{K_p}\exp\left(\mathrm{cos}(\bm{i}_{e_i}, \bm{t}_{e_i,k'}^{(p)})/\mu\right)}.
\end{equation}
For the $K_p$ class-specific words in the $p$-th aspect, the label prediction error for the traffic scene image $\mathcal{I}$ is estimated using a cross entropy loss function as
\begin{equation}~\label{eq:loss}
    \mathcal{L}^{(p)}_{\mathcal{I}} = - \sum_{k=1}^{K_p} {y}_{\mathcal{I},k}^{(p)} \log\left(\hat{y}_{\mathcal{I},k}^{(p)}\right),
\end{equation}
where ${y}_{\mathcal{I},k}^{(p)}=1$ if the image $\mathcal{I}$ corresponds to the traffic scene class $k$ in the aspect $p$; otherwise, we set it as 0.
For all training samples $\mathcal{I}$ and $P$ aspects, the final loss function is defined as
\begin{equation}\label{eq:total_loss}
    Loss = \sum_{\mathcal{I}} \sum^{P}_{p=1} \mathcal{L}^{(p)}_{\mathcal{I}}.
\end{equation}
The detailed training algorithm and the implementation details of the \name model are provided in the Appendix.

\vspace{-0.15cm}
\paratitle{Description Generation.} The \name model adopts a template-based approach for generating scene descriptions. Given a traffic scene image and its corresponding vehicle trajectories, we can generate an image feature $\bm{i}_{e_i}$ using the image encoder of CLIP and generate prompt feature vectors $\bm{t}_{e_i,k}^{(p)}$ using the SCAMP module. 
Next, we bring the image and prompt feature vectors
into Eq.~\eqref{eq:inference} to obtain the predicted class-specific words of different aspects for the image, \ie $[\mathrm{CLASS}]_{k*}^{(p)}$ with $p\in \{1,\ldots, P\}$. 
Then, we design a scene description template with placeholders for these class-specific words, such as \textit{The road is in the $[\mathrm{CLASS}]^{(1)}$ scene. The surface is $[\mathrm{CLASS}]^{(2)}$ and the width is $[\mathrm{CLASS}]^{(3)}$. It is $[\mathrm{CLASS}]^{(4)}$ to pass through.} Finally, we use the predicted class-specific words to replace these placeholders to generate the output traffic scene description of the \name model, as shown in Fig.~\ref{illustration}.
It is worth noting that the words in the template are not prompts and we don't input them into the \name model. The performance of predicting class-specific words is solely determined by the learnable prompts generated by the SCAMP module, and has no connection with the words in the template. Therefore, we do not have to take pains to design the template words as designing a hand-crafted prompt.

\vspace{-0.2cm}
\section{EXPERIMENTS}~\label{sec:experiments}
\vspace{-0.4cm}

\begin{table*}
    \footnotesize
    \centering
    \caption{Performance comparison on the Beijing dataset. All the results are better with larger values. The optimal results are shown in bold and the second-best results in the baseline models are underlined.}
    \begin{tabular}{c|@{\hspace{5pt}}c@{\hspace{5pt}}c@{\hspace{5pt}}c@{\hspace{5pt}}c@{\hspace{5pt}}c@{\hspace{5pt}}c@{\hspace{5pt}}c@{\hspace{5pt}}c}
    \toprule
    Dataset & \multicolumn{8}{c@{\hspace{5pt}}}{Beijing} \\
    \midrule
    Aspect & \multicolumn{2}{c@{\hspace{5pt}}}{Scene} & \multicolumn{2}{c@{\hspace{5pt}}}{Surface} & \multicolumn{2}{c@{\hspace{5pt}}}{Width} & \multicolumn{2}{c@{\hspace{5pt}}}{Accessibility} \\
    \midrule
    Metrics & ACC $\uparrow$ & F1 $\uparrow$ & ACC $\uparrow$ & F1 $\uparrow$ & ACC $\uparrow$ & F1 $\uparrow$ & ACC $\uparrow$ & F1 $\uparrow$ \\
    \midrule
    ResNet-50 & 0.487$\pm$0.023 & 0.410$\pm$0.035 & 0.831$\pm$0.011 & 0.447$\pm$0.027 & 0.336$\pm$0.032 & 0.285$\pm$0.018 & 0.638$\pm$0.039 & 0.314$\pm$0.031 \\
    ViT-B/32 & 0.490$\pm$0.029 & 0.415$\pm$0.017 & 0.824$\pm$0.008 & 0.441$\pm$0.011 & 0.350$\pm$0.028 & 0.290$\pm$0.014 & 0.645$\pm$0.032 & 0.351$\pm$0.025 \\
    \midrule
    
    RN50+GAT & 0.475$\pm$0.024 & 0.399$\pm$0.017 & 0.814$\pm$0.029 & 0.420$\pm$0.014 & 0.304$\pm$0.021 & 0.251$\pm$0.022 & 0.620$\pm$0.025 & 0.289$\pm$0.017 \\
    RN50+LSTM & 0.468$\pm$0.015 & 0.407$\pm$0.018 & 0.822$\pm$0.025 & 0.432$\pm$0.020 & 0.307$\pm$0.028 & 0.259$\pm$0.033 & 0.633$\pm$0.015 & 0.295$\pm$0.023 \\

    \midrule
    CLIP$_{\mathrm{ZS}}$ & 0.434 & 0.312 & 0.801 & 0.425 & 0.330 & 0.268 & 0.633 & 0.298 \\
    CoOp & 0.654$\pm$0.033 & \underline{0.578$\pm$0.029} & 0.721$\pm$0.024 & 0.452$\pm$0.018 & 0.548$\pm$0.013 & \underline{0.523$\pm$0.012} & 0.620$\pm$0.021 & 0.485$\pm$0.029 \\
    CoCoOp & 0.583$\pm$0.015 & 0.508$\pm$0.021 & 0.662$\pm$0.032 & 0.397$\pm$0.028 & 0.498$\pm$0.042 & 0.458$\pm$0.035 & 0.651$\pm$0.013 & 0.443$\pm$0.026 \\
    CLIP-Adapter & 0.440$\pm$0.026 & 0.351$\pm$0.019 & 0.790$\pm$0.020 & 0.431$\pm$0.027 & 0.336$\pm$0.014 & 0.285$\pm$0.011 & 0.622$\pm$0.024 & 0.314$\pm$0.016 \\
    Tip-Adapter & 0.589$\pm$0.034 & 0.530$\pm$0.027 & 0.804$\pm$0.012 & 0.433$\pm$0.018 & 0.557$\pm$0.024 & 0.486$\pm$0.033 & 0.725$\pm$0.013 & 0.515$\pm$0.020 \\
    Tip-Adapter-F & \underline{0.671$\pm$0.024} & 0.571$\pm$0.021 & \underline{0.849$\pm$0.009} & \underline{0.480$\pm$0.013} & \underline{0.571$\pm$0.032} & 0.488$\pm$0.028 & \underline{0.732$\pm$0.021} & \underline{0.522$\pm$0.019} \\
    ST-CLIP & \textbf{0.758$\pm$0.021} & \textbf{0.697$\pm$0.017} & \textbf{0.857$\pm$0.010} & \textbf{0.488$\pm$0.008} & \textbf{0.598$\pm$0.022} & \textbf{0.551$\pm$0.027} & \textbf{0.802$\pm$0.025} & \textbf{0.580$\pm$0.026} \\
    \midrule
    Improved & 13.1\% & 20.6\% & 0.9\% & 1.7\% & 4.7\% & 5.3\% & 9.6\% & 11.1\% \\
    \bottomrule
    \end{tabular}
    \label{tab:beijing_result}
\end{table*}

\begin{table*}
    \footnotesize
    \centering
    \caption{Performance comparison on the Chengdu dataset. All the results are better with larger values. The optimal results are shown in bold and the second-best results in the baseline models are underlined.}
    \begin{tabular}{c|@{\hspace{5pt}}c@{\hspace{5pt}}c@{\hspace{5pt}}c@{\hspace{5pt}}c@{\hspace{5pt}}c@{\hspace{5pt}}c@{\hspace{5pt}}c@{\hspace{5pt}}c@{\hspace{5pt}}}
    \toprule
    Dataset & \multicolumn{8}{c@{\hspace{5pt}}}{Chengdu} \\
    \midrule
    Aspect & \multicolumn{2}{c@{\hspace{5pt}}}{Scene} & \multicolumn{2}{c@{\hspace{5pt}}}{Surface} & \multicolumn{2}{c@{\hspace{5pt}}}{Width} & \multicolumn{2}{c@{\hspace{5pt}}}{Accessibility} \\
    \midrule
    Metrics & ACC $\uparrow$ & F1 $\uparrow$ & ACC $\uparrow$ & F1 $\uparrow$ & ACC $\uparrow$ & F1 $\uparrow$ & ACC $\uparrow$ & F1 $\uparrow$ \\
    \midrule
    ResNet-50 & 0.447$\pm$0.034 & 0.320$\pm$0.021 & 0.726$\pm$0.013 & 0.345$\pm$0.027 & 0.387$\pm$0.035 & 0.243$\pm$0.028 & 0.451$\pm$0.026 & 0.269$\pm$0.021 \\
    ViT-B/32 & 0.455$\pm$0.025 & 0.338$\pm$0.013 & 0.718$\pm$0.020 & 0.340$\pm$0.029 & 0.395$\pm$0.027 & 0.258$\pm$0.024 & 0.460$\pm$0.018 & 0.281$\pm$0.026 \\
    \midrule
    RN50+GAT & 0.415$\pm$0.019 & 0.285$\pm$0.027 & 0.701$\pm$0.016 & 0.302$\pm$0.023 & 0.346$\pm$0.032 & 0.205$\pm$0.024 & 0.389$\pm$0.017 & 0.224$\pm$0.015 \\
    RN50+LSTM & 0.428$\pm$0.026 & 0.312$\pm$0.021 & 0.702$\pm$0.011 & 0.305$\pm$0.014 & 0.367$\pm$0.019 & 0.231$\pm$0.020 & 0.432$\pm$0.019 & 0.265$\pm$0.028 \\
    \midrule
    CLIP$_{\mathrm{ZS}}$ & 0.435 & 0.281 & 0.710 & 0.341 & 0.276 & 0.225 & 0.433 & 0.235 \\
    CoOp & 0.689$\pm$0.012 & \underline{0.668$\pm$0.027} & 0.687$\pm$0.025 & \underline{0.424$\pm$0.031} & 0.457$\pm$0.028 & 0.418$\pm$0.021 & 0.659$\pm$0.016 & 0.412$\pm$0.020 \\
    CoCoOp & 0.609$\pm$0.021 & 0.578$\pm$0.020 & 0.630$\pm$0.014 & 0.392$\pm$0.023 & 0.387$\pm$0.025 & 0.363$\pm$0.026 & 0.521$\pm$0.018 & 0.328$\pm$0.024 \\
    CLIP-Adapter & 0.429$\pm$0.027 & 0.317$\pm$0.024 & 0.704$\pm$0.019 & 0.336$\pm$0.023 & 0.279$\pm$0.013 & 0.233$\pm$0.016 & 0.420$\pm$0.032 & 0.232$\pm$0.029 \\
    Tip-Adapter & 0.541$\pm$0.013 & 0.511$\pm$0.021 & 0.710$\pm$0.019 & 0.341$\pm$0.027 & 0.458$\pm$0.018 & 0.358$\pm$0.021 & 0.620$\pm$0.018 & 0.401$\pm$0.014 \\
    Tip-Adapter-F & \underline{0.694$\pm$0.012} & 0.630$\pm$0.020 & \underline{0.754$\pm$0.018} & 0.370$\pm$0.024 & \underline{0.501$\pm$0.017} & \underline{0.446$\pm$0.023} & \underline{0.737$\pm$0.024} & \underline{0.493$\pm$0.014} \\
    ST-CLIP & \textbf{0.779$\pm$0.025} & \textbf{0.685$\pm$0.016} & \textbf{0.788$\pm$0.024} & \textbf{0.450$\pm$0.018} & \textbf{0.521$\pm$0.017} & \textbf{0.467$\pm$0.022} & \textbf{0.843$\pm$0.018} & \textbf{0.529$\pm$0.027} \\
    \midrule
    Improved & 12.2\% & 2.5\% & 4.5\% & 6.1\% & 4.0\% & 4.7\% & 14.4\% & 7.3\% \\
    \bottomrule
    \end{tabular}
    \label{tab:chengdu_result}
\end{table*}

In this section, we conduct extensive experiments to demonstrate the effectiveness of our model.

\vspace{-0.4cm}
\subsection{Experimental Setup}

\subsubsection{Construction of the Datasets}

In the experiments, we utilize two real-world datasets consisting of traffic scene images and corresponding trajectories to evaluate the performance of our proposed model. Each data sample in these datasets comprises three components: a traffic scene image, a trajectory associated with the image, and class-specific word labels for different aspects of the traffic scene. 
The data was collected from the {\em DiDi-Rider} app platform, which is the largest online taxi-hailing and ride-sharing service app in China. In the datasets, the traffic scene images were captured by cameras mounted on vehicles, the taxi trajectories corresponding to these images were recorded by the app, and class-specific words were manually labelled by the DiDi company. The two datasets were collected in \emph{Beijing} and \emph{Chengdu} respectively, two metropolises in China with exceeding 20 million inhabitants. The road network data of these two cities was also collected for map matching using the FMM algorithm~\cite{yang2018fast}.
The detailed statistics are provided in the Appendix. 

\subsubsection{Methods to Compare}

In the experiment, we consider two types of baseline models for a comprehensive comparison.

\vspace{-0.1cm}
\paratitle{Visual Models.} These methods treat the TSU task as an image classification task, using pre-trained visual backbones to extract features from traffic scene images and training a simple classification head to predict labels for different aspects.

$\bullet$ \textit{ResNet-50}~\cite{he2016deep}: 
It is a deep convolutional neural network with 50 layers, designed for image classification tasks. It utilizes residual connections to mitigate the vanishing gradient problem, allowing for efficient training of very deep networks.

$\bullet$ \textit{ViT-B/32}~\cite{dosovitskiy2020image}: 
It is a Vision Transformer model that splits images into 32x32 patches and processes them using self-attention mechanisms. It achieves high performance in image classification by leveraging transformer architecture's ability to capture long-range dependencies.

\paratitle{Fusion-based Models.} 
These methods directly perform prediction by fusing image and spatio-temporal data (\emph{e.g.}, road networks and trajectories). They first extract features from each modality, then apply a late-fusion strategy to combine the multi-modal representations, and finally use a classification head for prediction.  

$\bullet$ \textit{ResNet-50+GAT}~\cite{velivckovic2017graph}:  
ResNet-50 is used to extract visual features from traffic scene images, while GAT encodes road network structures. The two feature representations are fused and fed into a classification head for prediction.  

$\bullet$ \textit{ResNet-50+LSTM}~\cite{hochreiter1997long}:  
ResNet-50 extracts image features, and LSTM models sequential dependencies from trajectory data. The fused representations are then passed through a classification head for prediction.  

\vspace{-0.1cm}
\paratitle{CLIP-based Extended Models.} These methods leverage the inherent knowledge of the pre-trained CLIP model by incorporating both visual and textual features. They utilize zero-shot or few-shot learning techniques to generate descriptive labels for various traffic scene images.

$\bullet$ \textit{CLIP}$_{ZS}$~\cite{radford2021learning}:
Since CLIP has inherent zero-shot transfer capability, we directly employ the CLIP model without fine-tuning for the TSU task. We follow the guideline of prompt engineering introduced by~\cite{radford2021learning}, adopting the hand-crafted prompts which is shown in the Appendix B.

$\bullet$ \textit{CLIP-Adapter}~\cite{gao2021clip}: 
It adds a learnable bottleneck layer finetuned on the training set. Although this improves flexibility, it still depends on static, hand-crafted prompts, limiting its ability to capture dynamic or contextual variations.

$\bullet$ \textit{Tip-Adapter}~\cite{zhang2021tip}: 
It caches training image features and class-specific word features, combining them with the input image feature at inference. This avoids backpropagation during inference, but the model heavily depends on cached representations and cannot adapt beyond the training distribution.

$\bullet$ \textit{Tip-Adapter-F}~\cite{zhang2021tip}: 
This model further fine-tunes the cache model of Tip-Adapter over the training set, which further boost its performance. Despite the improvement, its reliance on fixed prompts and cached features still constrains its generalization ability. Moreover, it ignores spatio-temporal contextual dynamics, which limits its applicability to complex traffic scenes.

$\bullet$ \textit{CoOp}~\cite{zhou2022learning}: 
It replaces hand-crafted prompts with learnable embeddings, with separate prompts learned for each aspect of traffic scene understanding. While more adaptive, the learned prompts remain limited to single-aspect information and fail to capture spatio-temporal contextual dependencies.

$\bullet$ \textit{CoCoOp}~\cite{zhou2022conditional}: 
It extends CoOp by introducing image-conditional tokens that enable dynamic prompts for different images. This improves generalization across domains, but it still inherits the same limitations as CoOp for TSU task.

\vspace{-0.3cm}
\subsection{Results and Analysis}

Table~\ref{tab:beijing_result} and Table~\ref{tab:chengdu_result} show the results of all comparison methods, where all methods except CLIP$_{\mathrm{ZS}}$ are trained with a training set of 16 few shots. 
We select ViT-B/32 as the vision backbone, which is widely adopted in the baselines~\cite{zhou2022learning, zhou2022conditional, zhang2021tip}. 
The performance is measured using the accuracy and macro F1 score for the class-specific word label classification, denoted as ``ACC'' and ``F1'' in the table. The performance are reported as the mean values and 95\% confidence intervals of five independent runs with varying random seeds.
From Table~\ref{tab:beijing_result} and Table~\ref{tab:chengdu_result}, we can observe the following results:

$\bullet$ Firstly, the CLIP model for zero-shot prediction still performs well, particularly in the surface classification task.
However, the performance improvement of visual methods fine-tuned solely on visual features remains limited. This is likely due to their heavy reliance on labeled data, which makes it difficult for such methods to quickly adapt to new tasks in a few-shot setting.
Moreover, fusion-based approaches perform even worse than visual methods. On the one hand, late fusion tends to undermine the discriminative power of pre-trained visual features; on the other hand, few-shot training struggles to learn effective cross-modal representations.

$\bullet$ Secondly, both CoOp and CoCoOp outperform CLIP$_{\mathrm{ZS}}$, indicating the effectiveness of learnable prompts. However, despite the addition of a lightweight neural network to generate for each image an input-conditional token, CoCoOp is not as effective as CoOp, indicating that a simple dynamic prompt cannot fully understand the dynamic traffic scenes. 
Additionally, the improvement achieved by CLIP-Adapter is also constrained, suggesting that the simple bottleneck design may discard fine-grained information, which is especially detrimental in high-variance tasks such as traffic scene understanding.

$\bullet$ Thirdly, Tip-Adapter outperforms CLIP$_{\mathrm{ZS}}$ despite without explicit training. 
It is attributed to the fact that although Tip-Adapter does not require training, it caches the training set as a part of model, essentially leveraging the knowledge of the training set. 
Moreover, by further refining the parameters of the cache model, Tip-Adapter-F achieves superior performance compared to all the baseline models.

$\bullet$ Finally, our proposed model, \name, consistently outperforms all the baselines, including both the visual models and CLIP-based extended models, across the Beijing and Chengdu datasets. \name explicitly incorporates ST-context for prompt learning, which is highly beneficial for TSU tasks. Moreover, the design of multiple prompts and the corresponding bi-level multi-aspect prompt attention mechanism facilitates both patch-wise and image-wise information sharing between different modalities and aspects, thereby enhancing performance in multi-aspect label classification.

\vspace{-0.3cm}
\subsection{Few-shot Experiments}

\begin{figure}
    \centering
    \includegraphics[width=1\linewidth]{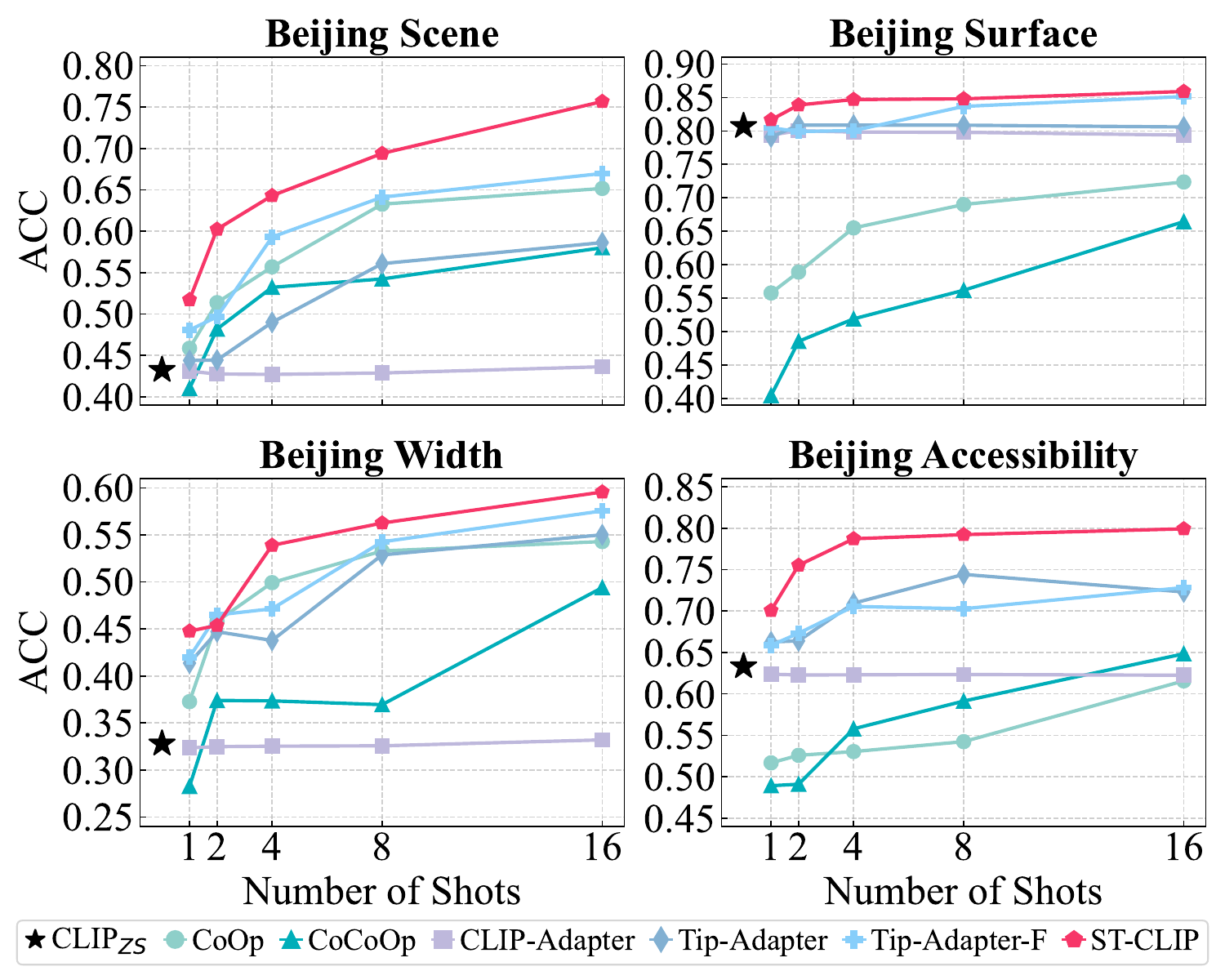}
    \caption{The results of few-shot learning on Beijing dataset. The x-axis represents the number of labeled training data.}
    \label{shots}
\end{figure}

Few-shot learning is a key feature of our model. To evaluate this capability, we experiment with 1, 2, 4, 8, and 16 training samples on the \emph{Beijing} dataset, as shown in Fig.~\ref{shots}.
The \emph{Chengdu} dataset yields similar trends and is thus omitted for brevity.
Figure~\ref{shots} highlights several key observations:

$\bullet$ Firstly, the performance of most models generally improves with an increasing number of samples, except for CLIP-Adapter. 
Moreover, CoOp outperforms CoCoOp in most cases.
This may be because the semantics of traffic scenes are relatively complex and cannot be fully captured by a simple adapter module, such as those used in CLIP-Adapter and CoCoOp, with few-shot learning.

$\bullet$ Secondly, Tip-Adapter and Tip-Adapter-F prove to be effective extensions of CLIP for few-shot learning. Moreover, with additional parameter updates, Tip-Adapter-F generally outperforms Tip-Adapter. Furthermore, as the number of training samples increases, the performance gap between them widens. Ultimately, Tip-Adapter-F achieves the best results in most cases compared to other baselines. 

$\bullet$ Finally, in most cases, the proposed \name model outperforms the other baselines. When the number of training samples exceeds 8, \name outperforms all the baseline models. It is noteworthy that, when trained on only 2 or 4 samples, \name achieves  performance comparable to or even superior to that of other baseline models trained on 16 samples. This result highlights the strong capability and robustness of our model in few-shot learning scenarios.

\vspace{-0.3cm}
\subsection{Ablation Study}

\begin{figure}
    \centering
    \includegraphics[width=1\linewidth]{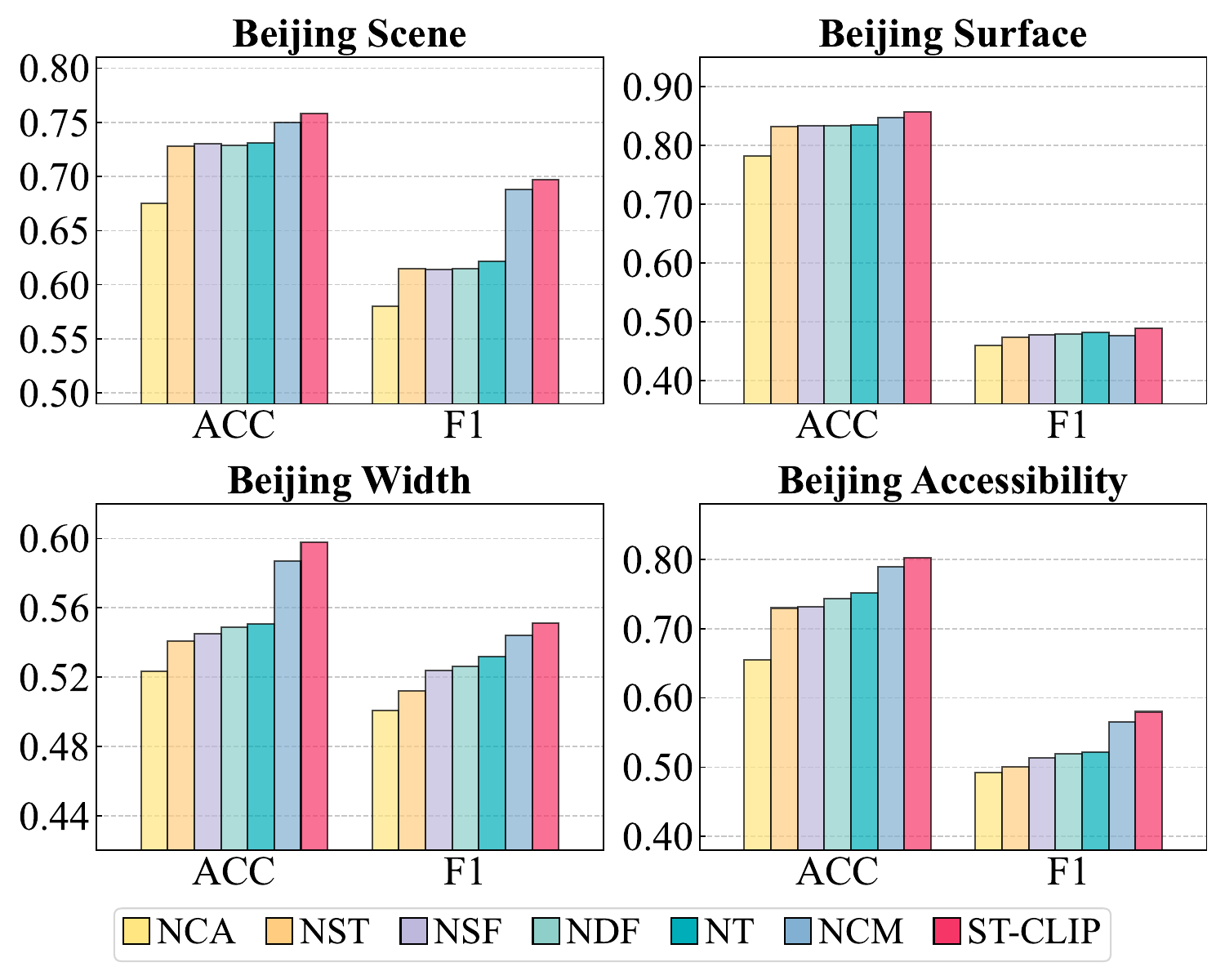}
    \caption{Ablation study on Beijing dataset for four tasks.}
    \label{ablation}
\end{figure}

In our model, we have incorporated both ST-context and a bi-level multi-prompt attention mechanism to enhance the performance of TSU tasks. Regarding the ST-context, we propose segment-level and trajectory-level ST-context. Regarding the bi-level multi-prompt attention mechanism, we introduce a patch-wise low-level cross-modal attention module and an image-wise high-level cross-aspect attention module. 
To investigate the individual contributions of each component to the overall performance, we conduct an ablation study on the \emph{Beijing} dataset. 
We prepare six variants of the proposed ST-CLIP model for comparisons, including 
$i$) \underline{NST}: It relies solely on the bi-level multi-prompt attention mechanism without ST-context, where $\tilde{\bm{v}}_{m,e_i}^{(p)} = \bm{w}_m^{(p)}$ in Eq.~\eqref{eq:context2prompt}. 
$ii$) \underline{NSF}: It removes the static features of the road segments, but only utilizes the dynamic properties as the initial features.
$iii$) \underline{NDF}: It omits the dynamic features of the road segments and retains the static properties.
$iv$) \underline{NT}: It only utilizes the segment-level context and the multi-prompt mechanism, but omits the trajectory encoder in Eq.~\eqref{eq:trajectory_encoder}, which provides context from neighbouring segments for the same trajectory. 
$v$) \underline{NCM}: It removes the low-level cross-modal attention mechanism while retaining the other modules.
$vi$) \underline{NCA}: It removes the high-level cross-aspect attention mechanism but retains the ST-context and the low-level attention module to generate learnable prompts. 

Fig.~\ref{ablation} presents the comparison results of the four tasks on \emph{Beijing} dataset. 
The performance of the complete \name model is also demonstrated. 
The performance rank can be roughly given as follows: \underline{NCA} $<$ \underline{NST} $<$ \underline{NSF} $<$ \underline{NDF} $<$ \underline{NT} $<$ \underline{NCM} $<$ ST-CLIP. 
Firstly, \underline{NCA} suffers the most performance degradation compared to \name, indicating the importance of the high-level cross-aspect attention mechanism. 
Besides, the performance gap between \underline{NST}, \underline{NSF} and \underline{NDF} demonstrates the effectiveness of ST-context information. 
Additionally, the comparison between \underline{NT} and \name shows that trajectory-level ST-context integration plays an auxiliary role in understanding traffic scenes. 
Moreover, the performance gap between \underline{NCM} and \name indicates the effect of the low-level cross-modal attention mechanism, which introduces the patch-wise visual features for learnable prompts.
Finally, \name achieves the best results in all tasks equipped with the complete modules, demonstrating the effectiveness of every module in our model.

\vspace{-0.3cm}
\subsection{Qualitative Analysis}
\begin{figure}[t]
    \centering
    \small
    \subfigure[Low-level Cross-modal Attention Heatmap]{
        \begin{minipage}[b]{0.95\linewidth}
        \includegraphics[width=1\linewidth]{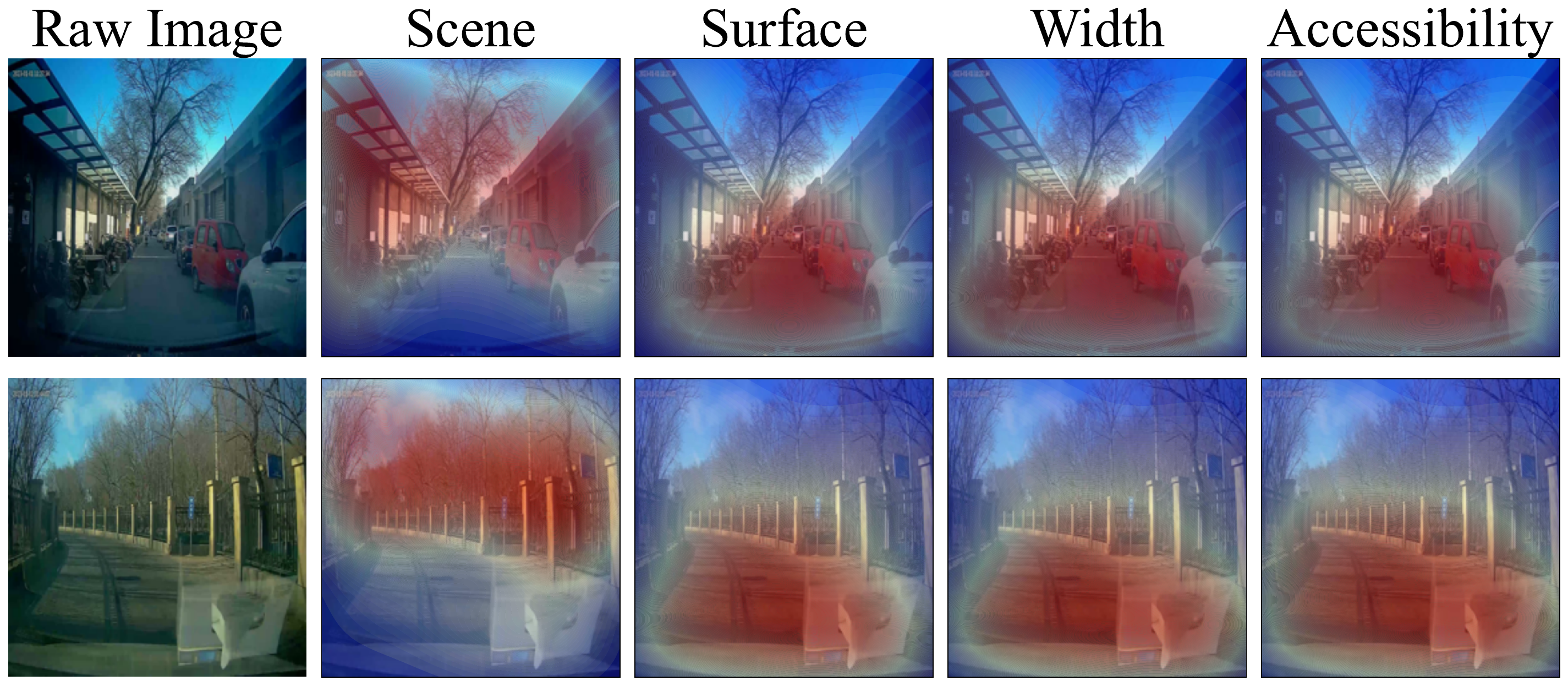}
        \vspace{-0.2cm}
        \label{fig:cross_modal_attention}
        \end{minipage} 
    }
    \subfigure[High-level Cross-aspect Attention Heatmap]{
        \begin{minipage}[b]{0.95\linewidth}
        \includegraphics[width=1\linewidth]{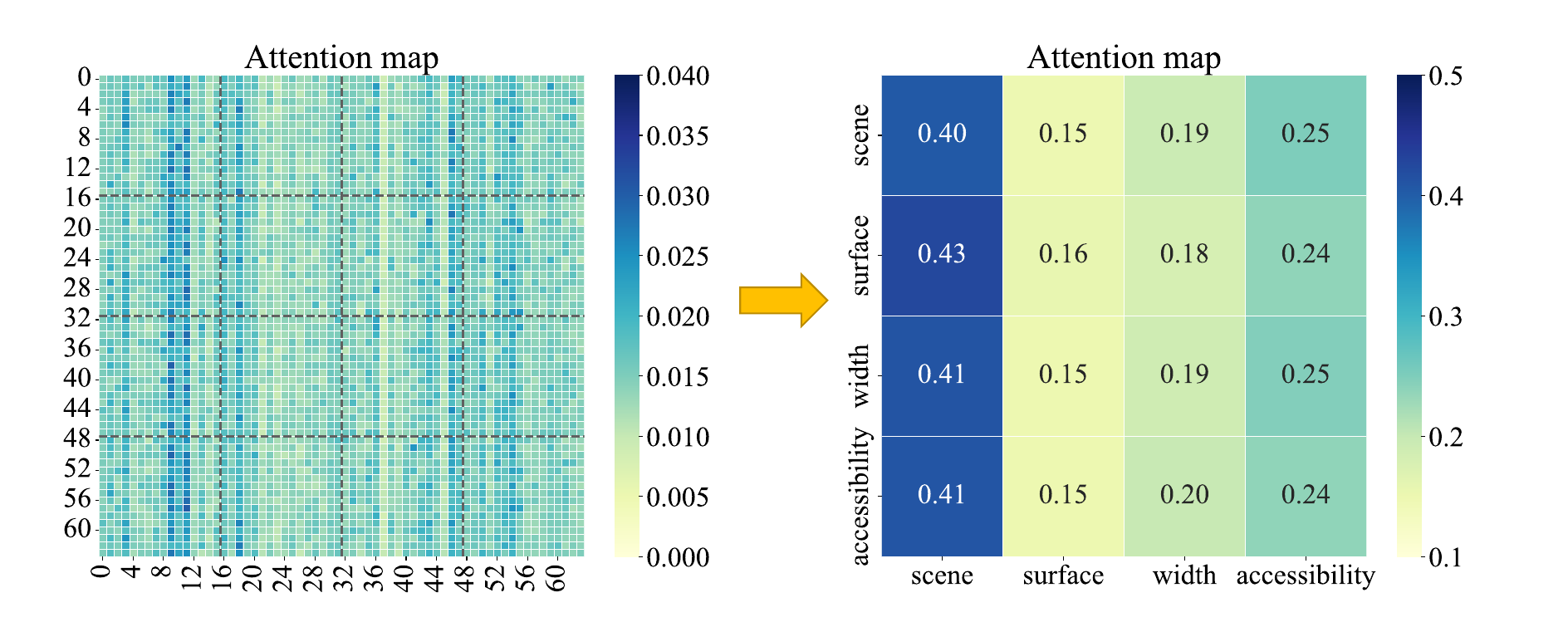}
        \vspace{-0.5cm}
        \label{fig:cross_aspect_attention}
        \end{minipage}
    }
    \caption{Bi-level multi-aspect prompt attention heatmap.}
    \label{attention_heatmap}
    \vspace{-10pt}
\end{figure}

\begin{figure}[t]
    \centering
    \small
    \subfigure[CoCoOp]{
    \begin{minipage}[b]{0.241\columnwidth}
    \includegraphics[width=1\linewidth]{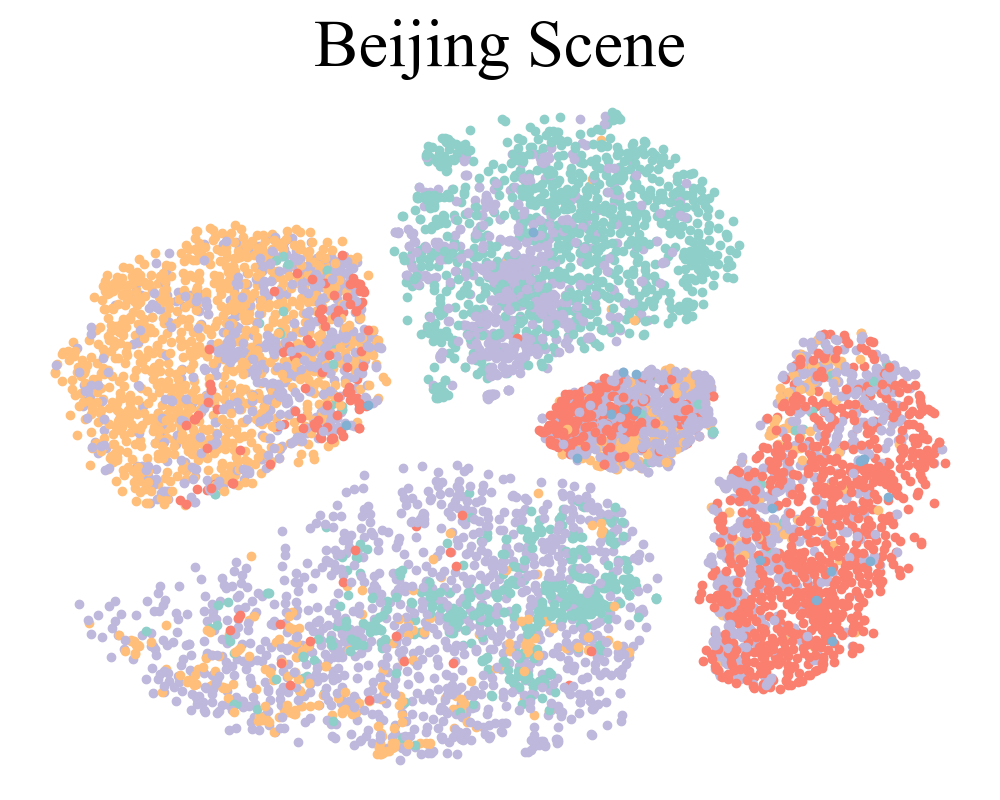}
    \end{minipage}
    \begin{minipage}[b]{0.241\columnwidth}
    \includegraphics[width=1\linewidth]{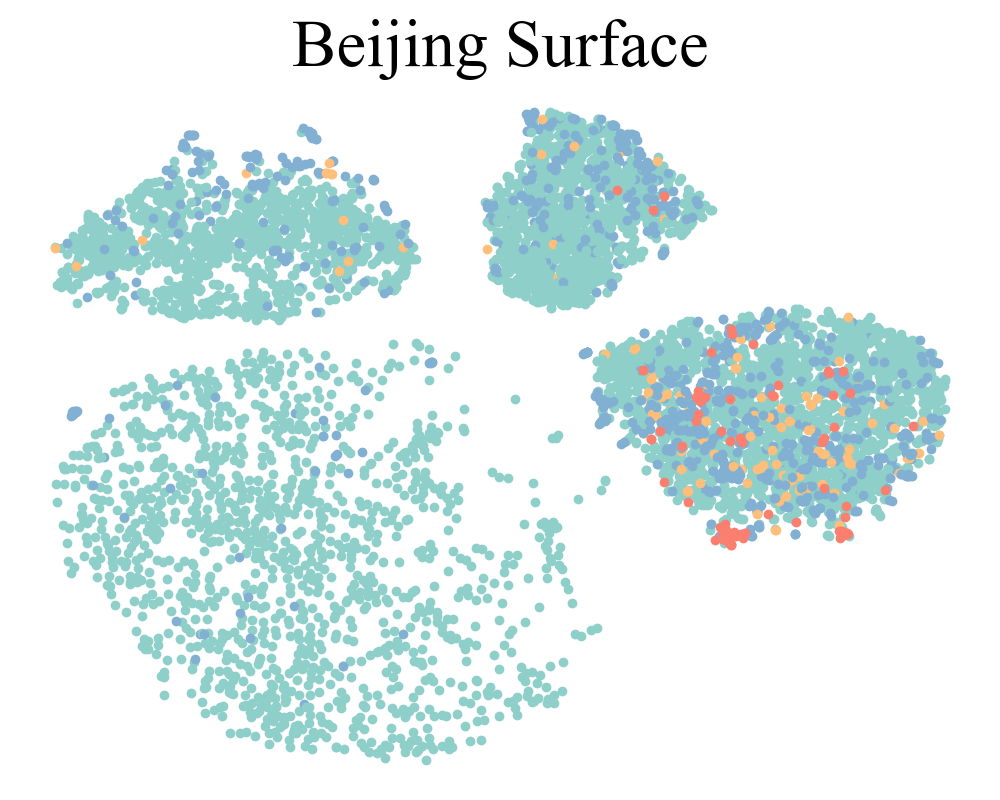}
    \end{minipage}
    \begin{minipage}[b]{0.241\columnwidth}
    \includegraphics[width=1\linewidth]{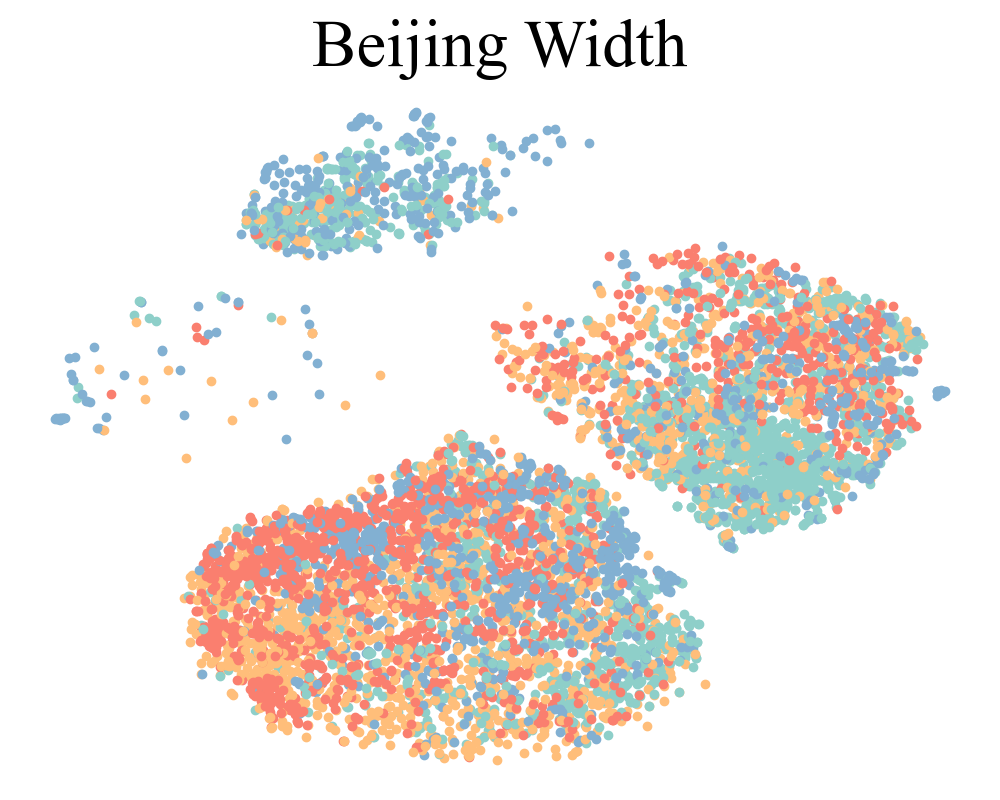}
    \end{minipage}
    \begin{minipage}[b]{0.241\columnwidth}
    \includegraphics[width=1\linewidth]{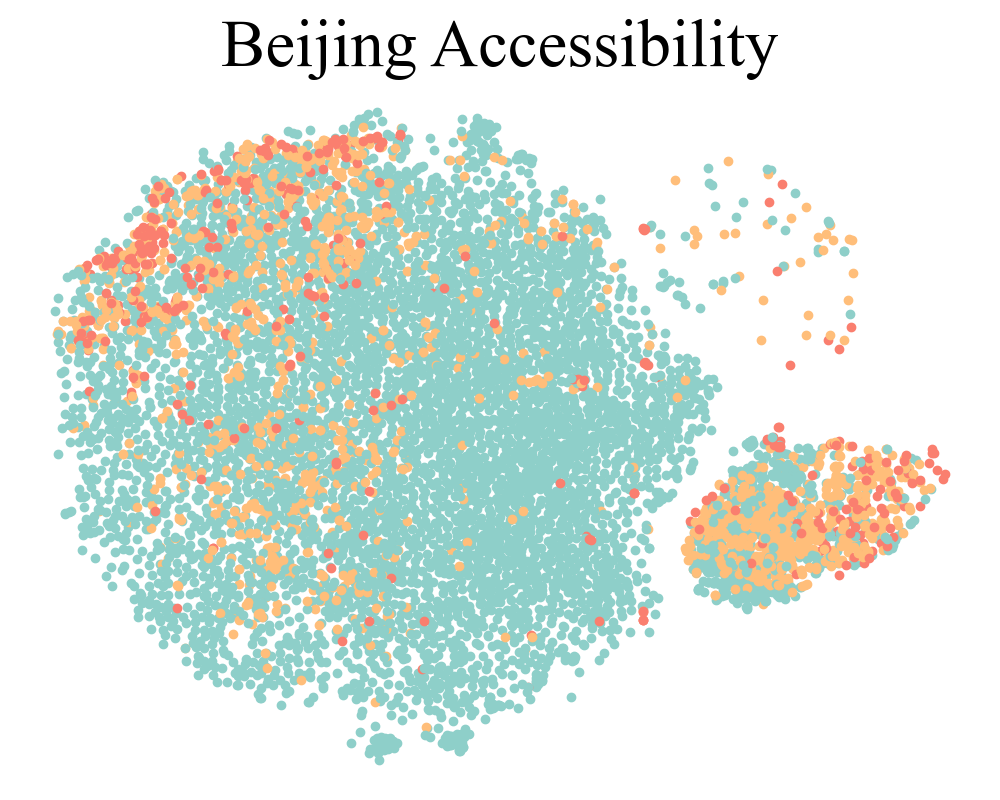}
    \end{minipage}
    \label{fig:tsne_CoCoOp}
    } \\
    \subfigure[ST-CLIP]{
    \begin{minipage}[b]{0.241\columnwidth}
    \includegraphics[width=1\linewidth]{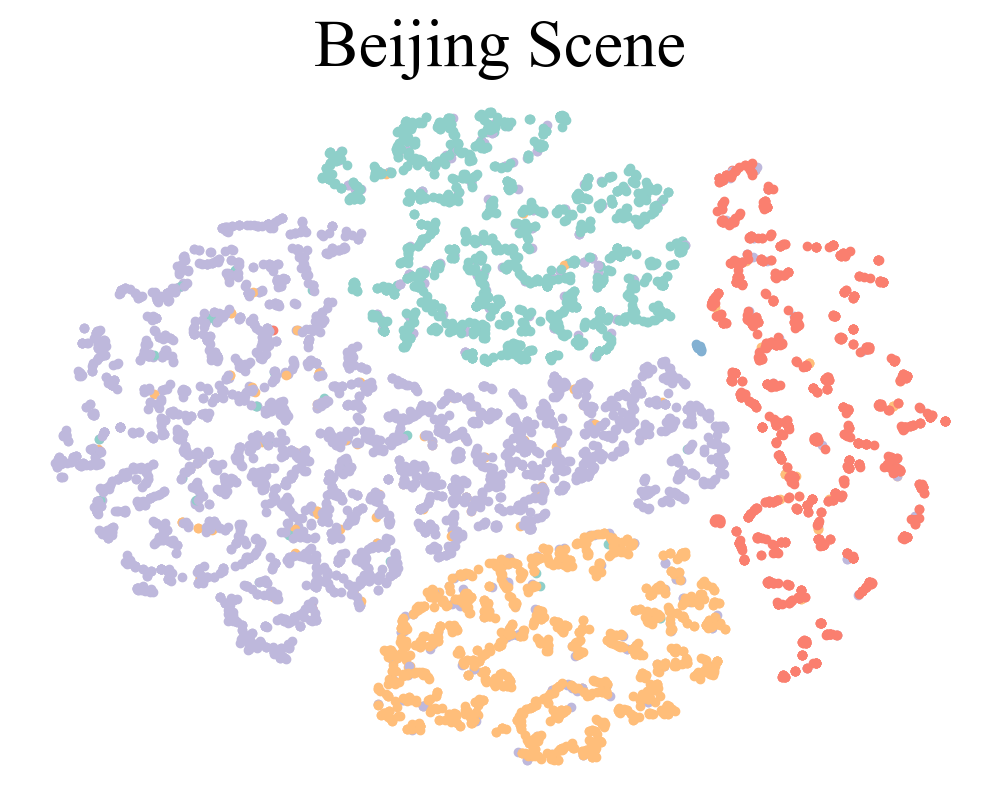}
    \end{minipage}
    \begin{minipage}[b]{0.241\columnwidth}
    \includegraphics[width=1\linewidth]{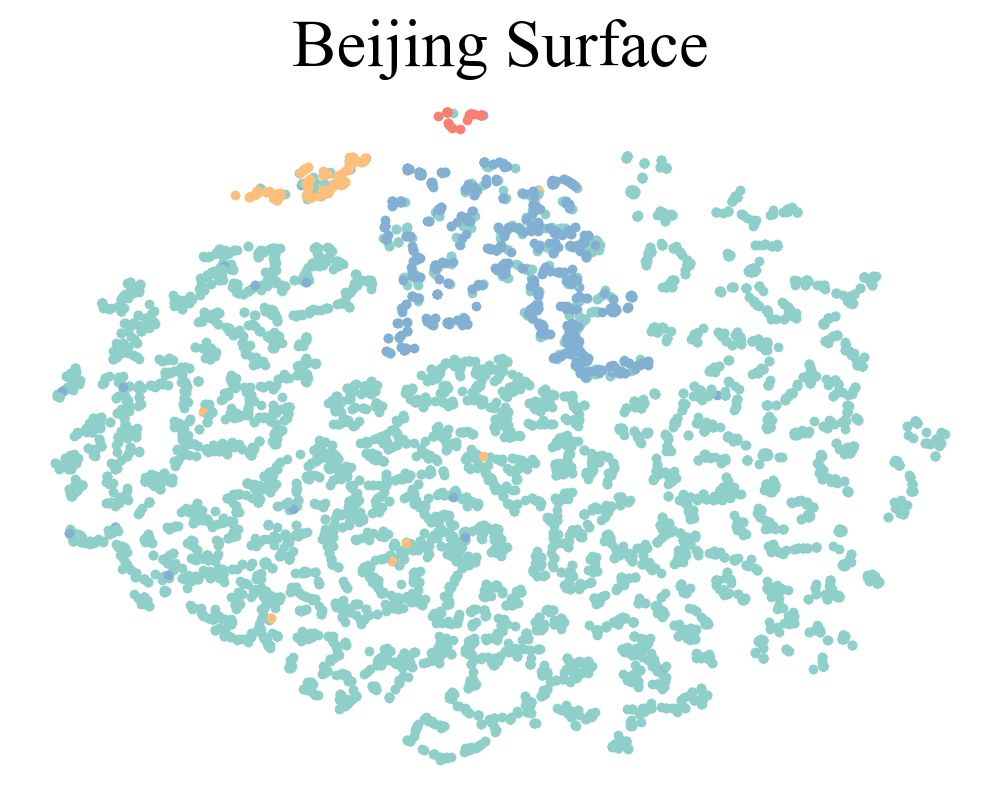}
    \end{minipage}
    \begin{minipage}[b]{0.241\columnwidth}
    \includegraphics[width=1\linewidth]{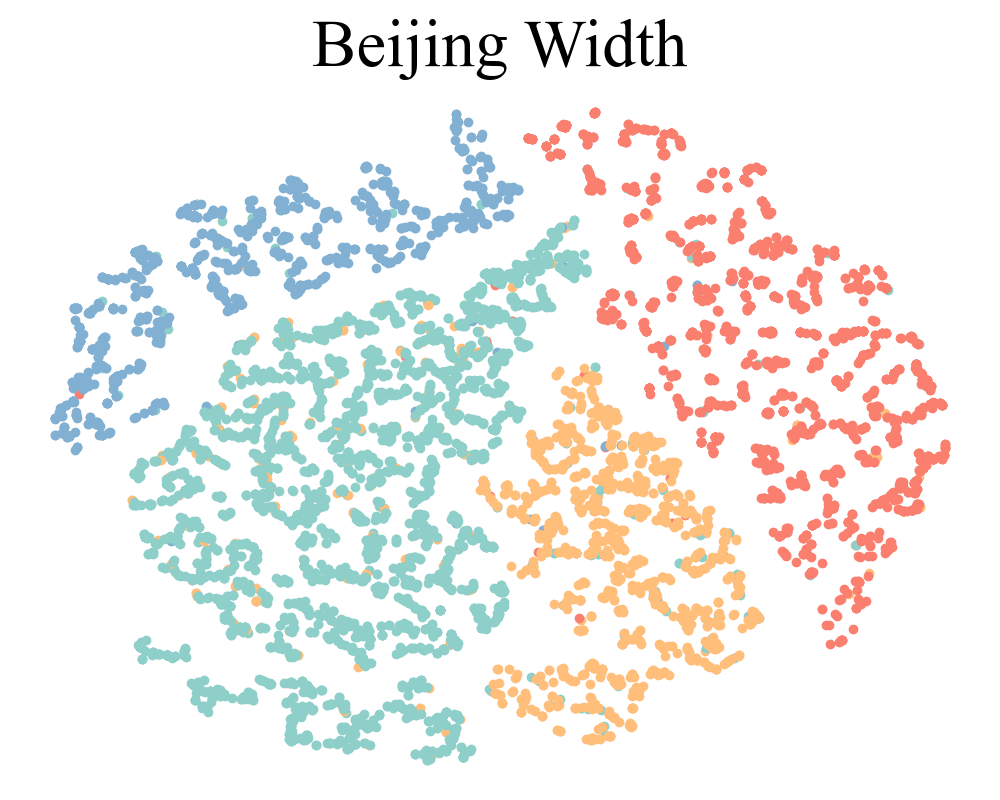}
    \end{minipage}
    \begin{minipage}[b]{0.241\columnwidth}
    \includegraphics[width=1\linewidth]{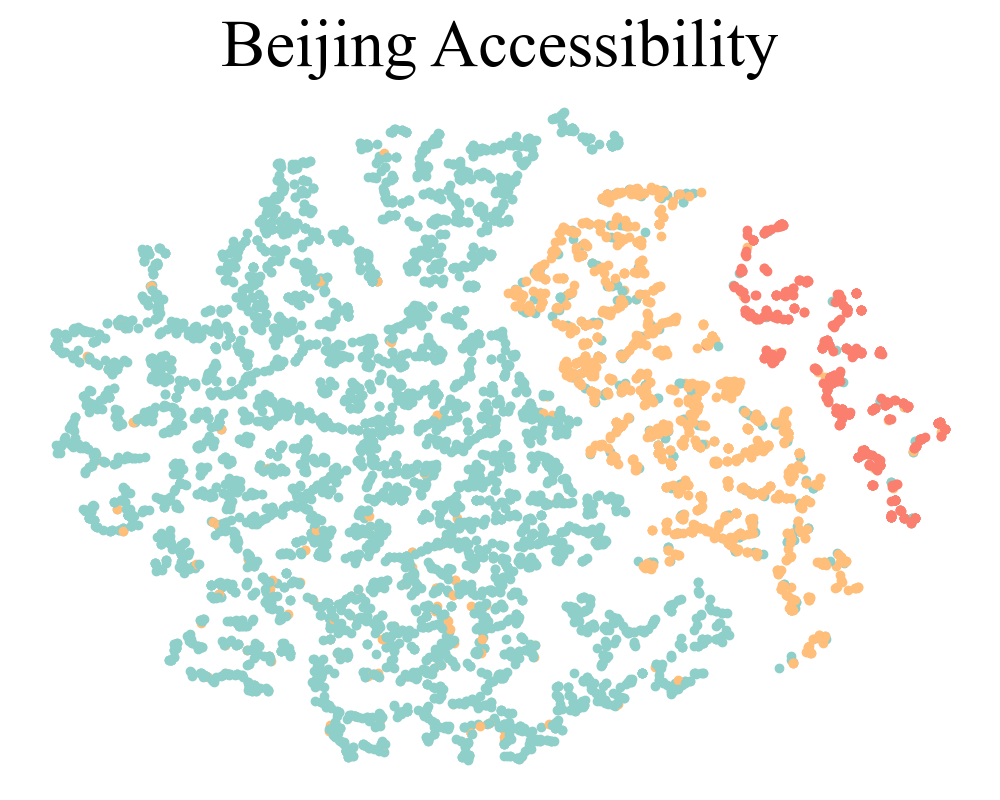}
    \end{minipage}
    \label{fig:tsne_ST-CLIP}
    }
    \caption{Visualization of text features on the Beijing dataset. The boundaries of features in ST-CLIP are clearly distinct, while in CoCoOp, different types of features are mixed together.}
    \label{tsne}
    \vspace{-10pt}
\end{figure}

\subsubsection{Bi-level Multi-aspect Prompt Attention Heatmap}

To illustrate the effectiveness of the bi-level multi-aspect prompt attention mechanism, we visualize the low-level and high-level attention heatmap respectively in Fig.~\ref{attention_heatmap}. 

Fig.~\ref{fig:cross_modal_attention} presents the kernel density estimation (KDE) of cross-modal attention for two examples, each showing a raw traffic scene image alongside the attention distribution of four different aspects across image patches.  
Colors indicate varying levels of attention, with red denoting high attention and blue denoting low attention.  
As illustrated, prompts corresponding to different aspects lead to distinct attention patterns. For instance, when analyzing the \emph{Scene} aspect, the model attends more to the upper background regions of the images, while for road-related aspects such as \emph{Width}, \emph{Surface}, and \emph{Accessibility}, the attention is concentrated on the lower half, typically corresponding to the road area. These observations confirm the effectiveness of our cross-modal attention mechanism in dynamically capturing aspect-specific visual cues.

Fig.~\ref{fig:cross_aspect_attention} presents the high-level cross-aspect attention heatmap. 
Considering that there are four aspects in total, each with a sequence length of $M=16$ learnable prompt feature vectors, the left part of Fig.~\ref{fig:cross_aspect_attention} illustrates the attention relationships among all $64$ learnable prompt word embedding vectors.
The heatmap values for each word pair are calculated as the average attention weights across all test samples. 
By summing the attention weights of each subregion and normalizing each row, the right part of Fig.~\ref{fig:cross_aspect_attention} reveals the attention heatmap across different aspects. 
We observe that the \emph{Scene} aspect is closely related to all aspects, as indicated by the darkest color in each row.
Additionally, the \emph{Accessibility} aspect also demonstrates a strong relationship with other aspects.
Notably, both \emph{Scene} and \emph{Accessibility} are the aspects where significant performance improvements are observed in Table~\ref{tab:beijing_result} and Table~\ref{tab:chengdu_result}. 
This attention relationship provides an explanation for the observed performance improvements. 
Conversely, the weak relationship of \emph{Surface} with other aspects may contribute to the limited performance improvement reported in Table~\ref{tab:beijing_result} and Table~\ref{tab:chengdu_result}. 

\subsubsection{Text Feature Visualization}

We visualize the feature vectors $\bm{t}^{(p)}_{e_i,k}$ of prompts generated by \name and CoCoOp over the \emph{Beijing} dataset using the t-SNE method~\cite{van2008visualizing}, as shown in Fig.~\ref{tsne}.
Among all the baselines, only CoCoOp produces distinct prompt feature vectors for different image inputs, whereas the other methods yield identical feature vectors for images with the same class-specific word. In the figure,
each dot represents a prompt feature vector, with different colors denoting different class-specific words for each aspect.

Fig.~\ref{fig:tsne_CoCoOp} illustrates the visualization results of the CoCoOp model across the four tasks, showing significant overlaps among different categories. Except for the \emph{Scene} dataset, where distinct clusters are visible, the other aspect datasets exhibit a mix of classes without clear separation. This lack of clear boundaries may explain the poor classification performance of CoCoOp. In contrast, Fig.~\ref{fig:tsne_ST-CLIP} demonstrates the \name model's performance, where clear boundaries are observed among different classes across all four tasks. Despite variations in sample numbers across different classes in some tasks, our model effectively captures the distinct characteristics of each class. This capability likely contributes to the superior performance of \name compared to CoCoOp.

\subsubsection{Case Study}

In this section, we present several case studies to illustrate our model's ability to generate descriptions for various traffic scene images. For comparison, we also use the top-performing baseline, Tip-Adapter-F, to generate descriptions for the same cases. For more case analyses, please refer to Appendix~F.

\begin{figure}[t]
    \centering
    \includegraphics[width=1\linewidth]{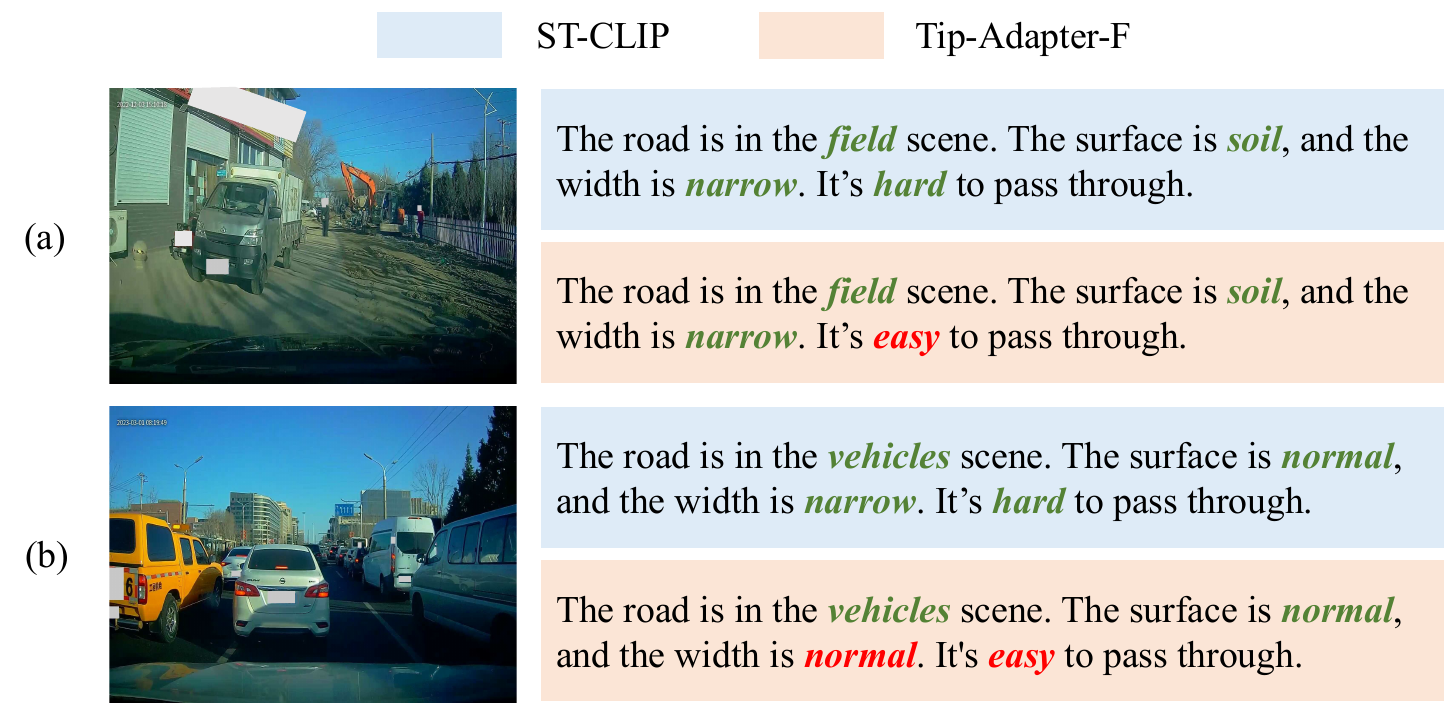}
    \caption{Case study of traffic scene descriptions generated by ST-CLIP and Tip-Adapter-F where green words represent the correctly classified keywords, and red words represent the misclassified keywords.}
    \vspace{-10pt}
    \label{case-study}
\end{figure}

The generated results are shown in Fig.~\ref{case-study}, where black words represent the predefined template, green words indicate correctly classified keywords, and red words highlight misclassified keywords. 
The results reveal that the descriptions generated by Tip-Adapter exhibit inconsistencies with both the images and the context. 
In contrast, \name provides more accurate and consistent descriptions. 
For instance, the road in Fig.~\ref{case-study}(a) is rough and narrow with many obstacles, making it difficult to pass through. \name accurately predicts this scenario, while Tip-Adapter-F correctly identifies the scene, surface, and width of the road, but fails to accurately predict the accessibility. This discrepancy may be attributed to Tip-Adapter-F's lack of consideration for the correlations between different aspects. Our \name model addresses this limitation through its bi-level multi-aspect prompt attention mechanism.
Additionally, in Fig.~\ref{case-study}(b), although the road is flat, heavy traffic during rush hour makes it challenging to pass through. Our model, by incorporating the characteristics of road network and trajectory behaviors in the ST-context, predicts more accurately compared to Tip-Adapter-F.

\vspace{-0.3cm}
\subsection{Time Analysis}

\begin{figure}[t]
    \centering
    \includegraphics[width=1\linewidth]{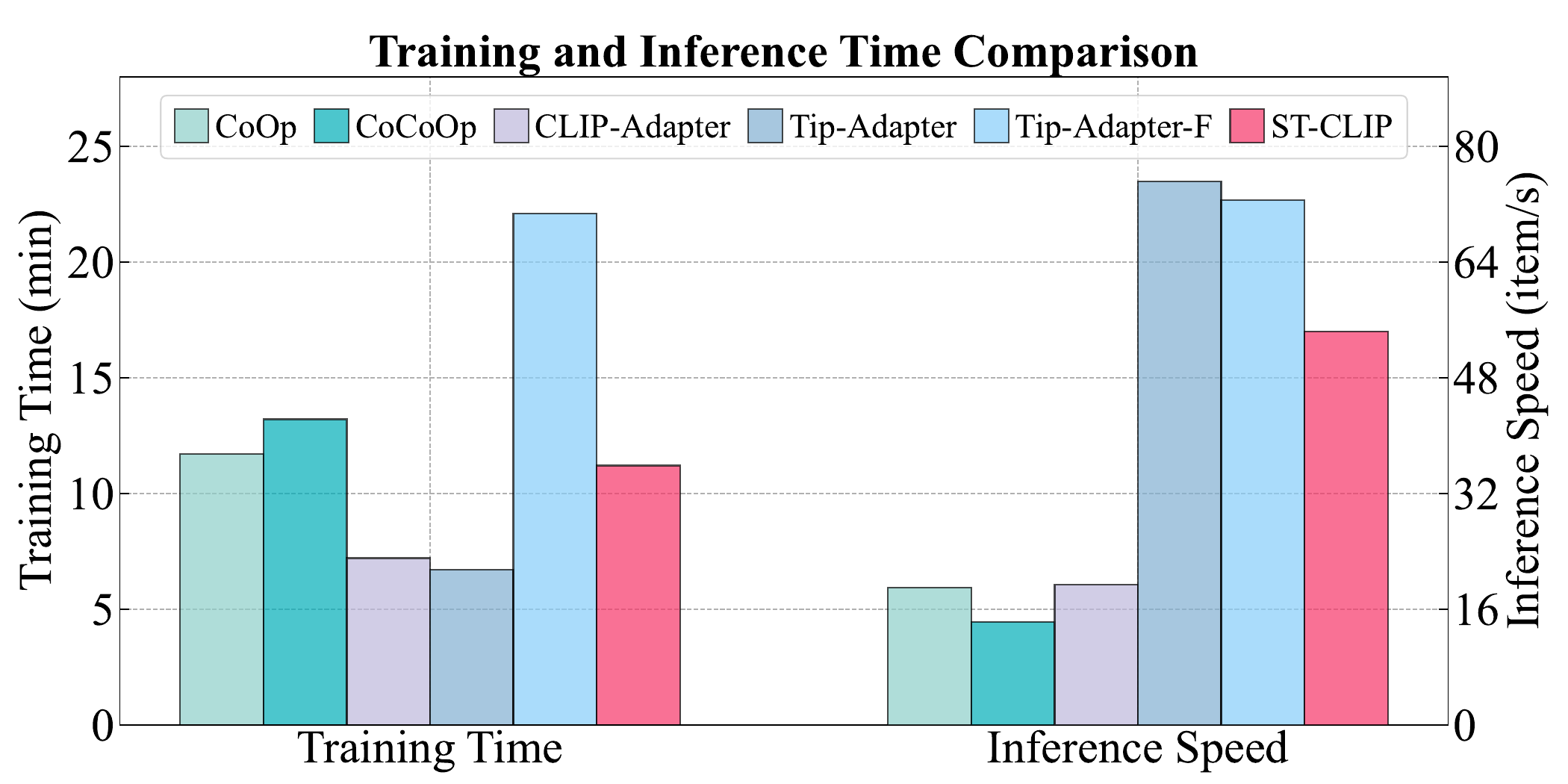}
    \caption{Time analysis of \name compared to baseline models.}
    \label{fig:time_analysis}
    \vspace{-10pt}
\end{figure}

To further evaluate the efficiency of our approach, we conducted experiments to compare our model with other baseline models in terms of training time and inference speed, as shown in Fig.~\ref{fig:time_analysis}.
The training time of \name is moderate. 
Compared to the optimal baseline model, Tip-Adapter-F, \name not only achieves superior experimental results but also significantly reduces training time, making it a more efficient and effective solution. This improvement can be attributed to several key factors.
First, Tip-Adapter-F relies on parameter initialization using features extracted from the training set, which introduces an inherent dependency on the dataset distribution. While this enhances model adaptation, it also adds computational overhead. Additionally, Tip-Adapter-F requires an extra hyperparameter search step after training to fine-tune its performance, further increasing training complexity and time consumption.
Regarding inference speed, \name constructs unique text features for each input image, necessitating the use of text encoder for every image. 
In contrast, Tip-Adapter-F uses fixed text features for all images, which are precomputed and reused, leading to faster inference speeds.
Compared to other baseline models, ST-CLIP achieves faster inference speed due to its ability to simultaneously predict all aspects without requiring one-by-one processing.

\vspace{-0.1cm}
\section{CONCLUSIONS}~\label{sec:coclusions}
\vspace{-0.3cm}

In this paper, we focus on bridging the gap between spatio-temporal data and visual-language models to address the challenge of traffic scene understanding. We propose a novel spatio-temporal enhanced model based on CLIP, a leading vision-language model in recent years. Our approach introduces a spatio-temporal context-aware multi-aspect prompt method to develop effective prompts for the backbone model, integrating spatio-temporal context with visual-textual data in the feature space. Extensive experimental results on two real-world datasets demonstrate the effectiveness and robustness of our proposed model. To our knowledge, this is the first attempt to integrate spatio-temporal knowledge into pre-trained multimodal models for traffic-related applications, shedding light on a novel research direction.

Our model adopts a discriminative approach, limiting generated traffic scene descriptions to predefined aspects and reducing expressiveness. A key limitation is the inability to produce fully context-aware narratives; future work will explore generative frameworks (e.g., LLMs) to enable more natural and comprehensive descriptions.
Moreover, the current approach also overlooks external factors such as environment and weather. Future research should address these by incorporating richer context, for instance using video clips instead of single-frame images.
Furthermore, we recognize the potential of combining spatio-temporal information with large-scale pre-trained models for traffic-related tasks.  Achieve a deeper understanding of the driving environment will require incorporating more detailed spatio-temporal data, such as points of interest. We leave this for further exploration to contribute to the development of more comprehensive and semantically enriched traffic maps.
\section*{Acknowledgments}
This work was supported by the National Natural Science Foundation of China (No. 72171013, 72222022, 72242101), the Fundamental Research Funds for the Central
Universities (JKF-2025017226182), and the DiDi Gaia Collaborative Research Funds.
\appendices

\section{Dataset Statistics}\label{sec-dataset}
% \vspace{-0.1cm}
\begin{table}[htbp]
    \small
    \caption{Statistics of the two datasets after preprocessing.}
    \centering
    \begin{small}
    \begin{tabular}{c|c|cc}
    \toprule
    {\bf Aspects} & {\bf Class-specific Words} & {\bf Beijing} & {\bf Chengdu} \\
    \midrule
    \multirow{5}*{Scene}
    & \#field & 1376 & 1629 \\
    %\cline{2-4}
    & \#vehicles & 1360 & 1444 \\
    %\cline{2-4}
    & \#alley & 1105 & 160 \\
    %\cline{2-4}
    & \#stall & 20 & 749 \\
    %\cline{2-4}
    & \#unknown & 3603 & 4341 \\
    \midrule
    \multirow{4}*{Surface}
    & \#normal & 6240 & 6293 \\
    %\cline{2-4}
    & \#broken & 140 & 1000 \\
    %\cline{2-4}
    & \#soil & 84 & 21 \\
    %\cline{2-4}
    & \#unknown & 1000 & 1000 \\
    \midrule
    \multirow{4}*{Width}
    & \#normal & 2528 & 2887 \\
    %\cline{2-4}
    & \#narrow & 1766 & 1868 \\
    %\cline{2-4}
    & \#extremely narrow & 1722 & 1727 \\
    %\cline{2-4}
    & \#unknown & 1448 & 1841 \\
    \midrule
    \multirow{3}*{Accessibility}
    & \#easy & 5428 & 6734 \\
    %\cline{2-4}
    & \#hard & 1628 & 1528 \\
    %\cline{2-4}
    & \#extremely hard & 408 & 61 \\
    \midrule
    \multicolumn{2}{c|}{\#image \& \#trajectory} & 7464 & 8323 \\
    \bottomrule
%    \textbackslash{} & \#trajectory & 7672 & 8660 \\
%    \hline
    \end{tabular}
    \end{small}
    \label{statistics}
\end{table}

\begin{table*}[htbp]
    \centering
    \footnotesize
    \caption{Overview of dataset schema across different CSV files.}
    \begin{tabular}{m{4.3cm} m{4.5cm} m{6.7cm}}
        \toprule
        \textbf{CSV File} & \textbf{Key Columns} & \textbf{Description} \\
        \midrule
        \texttt{image\_dataset.csv} & 
        \begin{itemize}[leftmargin=*]
          \item image\_path  
          \item label\_name\_list (list of strings)  
          \item label\_index\_list (list of ints)  
        \end{itemize}
        & 
        Full dataset of traffic scene images. Each row links an image to its multi-aspect labels (e.g., scene, surface, width, accessibility). The label information is stored as lists. \\
        
        \midrule
        \texttt{segment\_profile.csv} & 
        \begin{itemize}[leftmargin=*]
          \item segment\_id  
          \item function\_class, 
          \item lane\_number, 
          \item speed\_class  
          \item road\_length, 
          \item out\_degree  
          \item trajectory\_count, 
          \item medium\_speed  
          \item other\_attrs\_json  
        \end{itemize}
        & 
        Road segment profile table. Includes both static attributes (e.g., function class, lane number, road length) and dynamic statistics (e.g., trajectory count, median speed). The JSON field preserves additional extensible attributes. \\
        
        \midrule
        \texttt{image\_to\_trajectory.csv} & 
        \begin{itemize}[leftmargin=*]
          \item image\_path  
          \item trajectory\_segments (list of segment IDs)  
          \item image\_to\_segment  
        \end{itemize}
        & 
        Mapping table between images and their road-segment trajectories. Each image is linked to a sequence of road segments (trajectory), and the specific segment where the image was captured is also provided. For CSV storage, the segment list is represented as a semicolon-separated string. \\
        \bottomrule
    \end{tabular}
    \label{tab:dataset_overview_schema}
\end{table*}

\begin{table*}[htbp]
    \centering
    \footnotesize
    \caption{Schema of \texttt{image\_dataset.csv}.}
    \begin{tabular}{llll}
        \toprule
        \textbf{Column} & \textbf{Type} & \textbf{Description} & \textbf{Example}\\
        \midrule
        image\_path & string & Path to the traffic scene image. & path/to/image.jpg\\
        label\_name\_list & list(string) & List of the class label name. & [``vehicles'', ``normal'', ``normal'', ``hard'']\\
        label\_index\_list & list(int) & List of the class label index. & [1, 0, 0, 1]\\
        \bottomrule
    \end{tabular}
    \label{tab:image_dataset_schema}
\end{table*}

\begin{table*}[htbp]
    \centering
    \footnotesize
    \caption{Schema of \texttt{segment\_profile.csv}.}
    \begin{tabular}{llll}
        \toprule
        \textbf{Column} & \textbf{Type} & \textbf{Description} & \textbf{Example}\\
        \midrule
        segment\_id & int & Unique identifier of the road segment. & 0 \\
        function\_class & int & Degree of the segment function. & 4 \\
        lane\_number & int & Number of lanes. & 2 \\
        speed\_class & int & Speed limit in km/h. & 60 \\
        road\_length & float & Segment length in meters. & 114.6 \\
        out\_degree & int & Number of downstream segments. & 3 \\
        \midrule
        trajectory\_count & int & Number of passing vehicles in a time window. & 23 \\
        medium\_speed & float & Medium passing speed of passing vehicles in a time window in km/h. & 34.3 \\
        \midrule
        other\_attrs\_json & string & JSON dump of remaining attributes (optional, for extensibility). & / \\
        \bottomrule
    \end{tabular}
    \label{tab:segment_profile_schema}
\end{table*}

\begin{table*}[htbp]
    \centering
    \footnotesize
    \caption{Schema of \texttt{image\_to\_trajectory.csv}.}
    \begin{tabular}{llll}
        \toprule
        \textbf{Column} & \textbf{Type} & \textbf{Description} & \textbf{Example}\\
        \midrule
        image\_path & string & Path to the traffic scene image. & path/to/image.jpg \\
        trajectory\_segments & list(int) & List of segment IDs along the trajectory. & [4323, 3451, 3361, 3312, 2453]\\
        image\_to\_segment & int & Segment ID where the image is captured. & 3361 \\
        \bottomrule
    \end{tabular}
    \label{tab:image_to_trajectory_schema}
\end{table*}

\vspace{-0.2cm}
The number of images and corresponding trajectories for \emph{Beijing} and \emph{Chengdu} exceeds 7,400 and 8,300 respectively, spanning from December 1, 2022 to February 1, 2023. The statistics of the traffic scene images in different aspects with class-specific word labels are listed in Table~\ref{statistics}, where ``vehicles'' indicates a driving scene heavily populated with vehicles and ``stall'' represents a scene with many vendors or shops.

In addition to the traffic scene image datasets, we utilize the road networks and corresponding taxi trajectories to calculate the segment-level ST-context representations for these two cities. The numbers of road segments for \emph{Beijing} and \emph{Chengdu} are 38,775 and 12,548, respectively. The trajectories used to calculate segment representations, which do not contain traffic scene images, amount to over 1.7 million and 0.9 million for these two cities, respectively. All of these datasets were collected from the DiDi-Rider app platform.

To improve transparency and reproducibility, we provide an \emph{overview table} (Table~\ref{tab:dataset_overview_schema}) summarizing the role of each dataset CSV file, their key columns, and their purposes. This table gives a quick reference of how the dataset is organized. Then, we offer a detailed description of each file (Table~\ref{tab:image_dataset_schema}, Table~\ref{tab:segment_profile_schema}, Table~\ref{tab:image_to_trajectory_schema}) to ensure clarity and reproducibility.  

\section{Implementation Details}\label{sec:implementation}

We conduct our experiments on a machine equipped with an Intel(R) Xeon(R) CPU E5-2630 v4 @ 2.20GHz, 256GB of RAM, and a NVIDIA Tesla P40 GPU with 12GB of VRAM. 
The operating system used is Ubuntu 20.04.4 LTS, and the programming language is Python 3.8.13.
Our model is implemented using the PyTorch 1.8.0 library, with all the experiments executed on a single GPU. 
The model is trained using the SGD optimizer, with a batch size of 32 and an initial learning rate of 0.002, which decays according to the cosine annealing rule.
The maximum number of epochs is set to 100 for 16/8 shots, 50 for 4/2 shots, and 20 for 1 shot, except for the surface dataset where the maximum epoch is fixed to 30. 
The window size in all the experiments is consistently set to 3. 
The learnable context vectors are initialized from a zero-mean Gaussian distribution with a standard deviation of 0.02. 
The model dimension is set to 512, and the temperature parameter is configured following the CLIP model's settings.

As different hand-crafted prompts may significantly impact the experimental results~\cite{zhou2022learning}, 
we follow the guidelines of prompt engineering and adopt the prompts for CLIP$_{ZS}$, CLIP-Adapter, Tip-Adapter and Tip-Adapter-F as follows: ``A photo of a car driving in the $[\mathrm{CLASS}]^{(1)}$ scene'' for the ``scene'' aspect, ``A photo of a car driving on the $[\mathrm{CLASS}]^{(2)}$ surface'' for the ``surface'' aspect, ``A photo of a car driving on the $[\mathrm{CLASS}]^{(3)}$ road'' for the ``width'' aspect, and ``A photo of a car driving on the road which is $[\mathrm{CLASS}]^{(4)}$ to pass through'' for the ``accessibility'' aspect, where $[\mathrm{CLASS}]^{(p)}$ is the placeholder of the class-specific word for each aspect.

\vspace{-0.1cm}
\section{Training Algorithm}\label{sec:algorithm}

\begin{algorithm}[htbp]
\small
\SetAlgoLined
\LinesNumbered
\KwIn{Traffic scene images $\mathcal{I}$, road network $\mathcal{G}$, vehicle gps-based trajectories, class-specific words [CLASS]$^{(p)}_k$, word labels $y$.}
\KwOut{The well-trained ST-CLIP  model.}
Convert gps-based trajectories to road segment-based trajectories $\bm{tr}$ with $\mathcal{G}$ by FMM~\cite{yang2018fast}\;
Initialize the learnable prompts $\bm{w}_{m}^{(p)}$\;
Freeze the parameters of the CLIP base model\;
\For{each epoch}{
    \For{each batch}{
        \tcp{Extract ST-context}
        Generate segment-level feature embedding $\bm{h}_{e_i}^{(0)}$ for each segment of $\bm{tr}$ by Eq. (8)--(10)\;
        Perform trajectory-level ST-context learning with $\bm{h}_{e_i}^{(0)}$ to obtain $\bm{r}_{e_i}$ by Eq.~(11)--(13)\;
        \tcp{Extract image features}
        Generate patch features $\bm{F}_p$ and image feature $\bm{i}_{e_i}$ by Eq.~(18)\;
        \tcp{Extract text features}
        Merge ST-context $\bm{r}_{e_i}$ and learnable prompts $\bm{w}_{m}^{(p)}$ to obtain $\tilde{\bm{V}}_{e_i}^{(p)}$ by Eq.~(16)--(17)\;
        Perform low-level cross-modal prompt attention with $\tilde{\bm{V}}_{e_i}^{(p)}$ and $\bm{F}_p$ to obtain $\hat{\bm{V}}_{e_i}^{(p)}$ by Eq.~(19)--(20) \;
        Perform high-level cross-aspect prompt attention with $\hat{\bm{V}}_{e_i}^{(p)}$ to obtain ${\bm{V}}_{e_i}^{(p)}$ by Eq.~(21)--(22)\;
        Concatenate $\bm{\mathcal{V}}^{(p)}_{e_i}$ and the embedding vector of [CLASS]$^{(p)}_k$ to obtain $\bm{T}^{(p)}_{e_i,k}$ by Eq.~(23)\;
        Generate text feature $\bm{t}^{(p)}_{e_i,k}$ with $\bm{T}^{(p)}_{e_i,k}$ by Eq.~(24)\;
        \tcp{Calculate loss}
        Calculate the loss function with $\bm{i}_{e_i}$, $\bm{t}^{(p)}_{e_i,k}$ and labels $y$ by Eq.~(25)--(27)\;
        \tcp{Back propagation}
        Update model parameters with back propagation.
    }
}
\caption{The training process of ST-CLIP}
\label{training-algorithm}
\end{algorithm}

Algorithm~\ref{training-algorithm} outlines the training process of ST-CLIP, with notations summarized in Table~\ref{tab:notations}.
In line 1, raw GPS-based trajectories are map-matched to road segments using Fast Map Matching (FMM). Lines 2–3 initialize the learnable prompts while keeping the CLIP backbone frozen. During each training epoch, the model iterates over mini-batches (lines 4–17): segment-level features and spatio-temporal context are extracted (lines 6–7); patch-level and global image features are obtained by ViT (line 8); and enriched prompt features are constructed by integrating spatio-temporal context with learnable prompts (line 9). Lines 10–11 apply a bi-level cross-modal attention mechanism to refine these prompts, which are then combined with class-specific textual tokens to generate the final text features (lines 12–13). The loss is computed (line 14) and parameters updated via backpropagation (line 15).
This iterative process enables ST-CLIP to jointly capture visual-textual and spatio-temporal relationships for traffic scene understanding.

\begin{table*}[htbp]
    \centering
    \small
    \caption{The Notations, explanations, and configurations in this work}
    \begin{tabular}{c|l|l|l}
    \toprule
        \textbf{Group} & \textbf{Notation} & \textbf{Explanation} & \textbf{Configuration} \\
    \midrule
        \multirow{11}{*}{\makecell[c]{Spatio-\\Temporal\\Data\\Related}}& $\bm{h}_{(\cdot)}\in\mathbb{R}^{D_p}$ & The embedding vector of road property $(\cdot)$. & $D_p=64$ \\
        \cline{2-4}
        & $\bm{h}_s^{(s)}\in\mathbb{R}^{D_p\times N_s}$ & \makecell[l]{The embedding vector of static properties for \\ segment $s$.} & \makecell[l]{$D_p=64$\\ $N_s=6$} \\
        \cline{2-4}
        & $\bm{h}_{e_i}^{(d)}\in\mathbb{R}^{D_p\times N_d}$ & \makecell[l]{The embedding vector of time-varying properties \\ for segment $s$ at time $\tau$.} & \makecell[l]{$D_p=64$\\ $N_d=2$} \\
        \cline{2-4}
        & $\bm{h}_{e_i}^{(0)}\in\mathbb{R}^{D}$ & The feature vector of segment $s$ at time $\tau$. & $D=512$ \\
        \cline{2-4}
        & $\bm{H}^{(0)}\in\mathbb{R}^{(2N_w+1)\times D}$ & The initial tracklet representation matrix. & \makecell[l]{$D=512$\\ $N_w=1$} \\
        \cline{2-4}
        & $\bm{H}^{(L)}\in\mathbb{R}^{(2N_w+1)\times D}$ & \makecell[l]{The final tracklet representation matrix after \\ Transformer.} & \makecell[l]{$D=512$\\ $N_w=1$} \\
        \cline{2-4}
        & $\bm{r}_{e_i}\in\mathbb{R}^D$ & The representation of trajectory sample $e_i$. & $D=512$ \\
    \midrule
        \multirow{22}*{\makecell[c]{Text\\Related}} & $\bm{\mathcal{V}}^{(p)}\in\mathbb{R}^{M\times D}$ & \makecell[l]{The word embeddings of static prompts for the \\ $p$-th aspect.} & \makecell[l]{$M=16$\\ $D=512$} \\
         \cline{2-4}
         & $\bm{\mathcal{W}}^{(p)}\in\mathbb{R}^{M\times D}$ & \makecell[l]{The features of learnable prompts for the $p$-th \\ aspect.} & \makecell[l]{$M=16$\\ $D=512$} \\
         \cline{2-4}
         & $\tilde{\bm{\mathcal{V}}}_{e_i}^{(p)}\in\mathbb{R}^{M\times D}$ & \makecell[l]{The features of ST-aware learnable prompts \\ for the $p$-th aspect.} & \makecell[l]{$M=16$\\ $D=512$} \\
         \cline{2-4}
         & \makecell[l]{$\bm{W}_q^{(p)}\in\mathbb{R}^{D\times D}$ \\ $\bm{W}_k^{(p)}\in\mathbb{R}^{D\times D}$ \\ $\bm{W}_v^{(p)}\in\mathbb{R}^{D\times D}$} & \makecell[l]{The learnable parameters in cross-modal attention \\ for the $p$-th aspect.} & $D=512$ \\
         \cline{2-4}
         & \makecell[l]{$\bm{Q}^{(p)}\in\mathbb{R}^{M\times D}$ \\ $\bm{K}^{(p)}\in\mathbb{R}^{M\times D}$ \\ $\bm{V}^{(p)}\in\mathbb{R}^{M\times D}$} & \makecell[l]{The query, key and value of cross-modal attention \\ for the $p$-th aspect.} & \makecell[l]{$M=16$\\ $D=512$} \\
         \cline{2-4}
         & $\hat{\bm{\mathcal{V}}}_{e_i}^{(p)}\in\mathbb{R}^{M\times D}$ & \makecell[l]{The features of cross-modal ST-aware prompts for \\ the $p$-th aspect.} & \makecell[l]{$M=16$\\ $D=512$} \\
         \cline{2-4}
         & $\bm{W}_A^{(pq)}\in\mathbb{R}^{D\times D}$ & \makecell[l]{The learnable parameter in cross-aspect attention \\ between the $p$-th and $q$-th aspects.} & $D=512$ \\
         \cline{2-4}
         & $\bm{\mathrm{ATT}}_{p,q}\in\mathbb{R}^{M\times M}$ & \makecell[l]{The attention matrix between different prompts\\ of aspects.} & $M=16$ \\
         \cline{2-4}
         & $\bm{\mathcal{V}}_{e_i}^{(p)}\in\mathbb{R}^{M\times D}$ & The final features of prompts for the $p$-th aspect. & \makecell[l]{$M=16$\\ $D=512$} \\
         \cline{2-4}
         & $\bm{c}_k^{(p)}\in\mathbb{R}^D$ & The word embedding of word [CLASS]$_k^{(p)}$. & $D=512$ \\
         \cline{2-4}
         & $\bm{T}^{(p)}_{e_i,k}\in\mathbb{R}^{(M+1)\times D} $ & \makecell[l]{The concatenation of final prompt features and the \\ word embedding of $k$-th class for the $p$-th aspect.} & \makecell[l]{$M=16$\\ $D=512$} \\
         \cline{2-4}
         & $\bm{t}^{(p)}_{e_i,k}\in\mathbb{R}^D$ & \makecell[l]{The final text feature for the $k$-th class in the \\ $p$-th aspect.} & $D=512$ \\
    \midrule
       \multirow{4}{*}{\makecell[c]{Image\\Related}} & $\mathcal{I}\in\mathbb{R}^{H\times W\times C}$ & The traffic scene image. & \makecell[l]{$H=1080$\\ $W=1920$\\ $C=3$} \\
         \cline{2-4}
         & $\bm{i}\in\mathbb{R}^D$ & The image feature vector. & $D=512$ \\
         \cline{2-4}
         & $\bm{F}_p\in\mathbb{R}^{N_p\times D}$ & The patch features of image. & \makecell[l]{$N_p=16$\\ $D=512$} \\
    \bottomrule
    \end{tabular}
    \label{tab:notations}
\end{table*}

The time complexity of the \name model mainly comes from two components: ST-context representation learning and the bi-level multi-aspect prompt attention mechanism.
For ST-context representation learning, the complexity is $\mathcal{O}(2N_w^2D + N_wD^2)$, where $N_w$ is the window size of tracklets and $D$ the feature dimension. Since $N_w$ is a small constant, this reduces to $\mathcal{O}(D^2)$.
For the bi-level multi-aspect prompt attention, the complexity consists of low-level cross-modal attention $\mathcal{O}(MPN_pD)$ and high-level cross-aspect attention $\mathcal{O}((MP)^2D)$, where $M$ is the prompt length, $P$ the number of aspects, and $N_p$ the number of patches. Thus, the overall complexity simplifies to $\mathcal{O}(D^2 + (MP)^2D)$, depending only on feature dimension, number of aspects, and prompt length, confirming the model’s efficiency.

To further assess model complexity, we profile the computational efficiency of ST-CLIP and compare it against representative baselines, including CoOp, CoCoOp, CLIP-Adapter, and Tip-Adapter-F.
Three standard metrics are reported:
(1) \textbf{FLOPs}: number of floating-point operations per forward pass. 
(2) \textbf{Inference Latency}: mean time for a forward pass, with standard  deviation across multiple runs.
(3) \textbf{Peak Memory}: maximum GPU memory consumption during inference.

Table~\ref{tab:efficiency_comparison} summarizes the results. Compared with CoOp, ST-CLIP increases inference latency from $13.4$ ms to $36.7$ ms ($\approx 2.7\times$), while the gap with Tip-Adapter-F ($28.2$ ms) is about $30\%$. Similar patterns are observed for FLOPs and memory usage.
Importantly, ST-CLIP produces predictions for multiple aspects within a single forward pass, whereas baseline methods require separate models for each aspect. Thus, although ST-CLIP exhibits moderately higher cost per model, the overall efficiency remains favorable—its overhead is far lower than the na\"{\i}ve $4\times$ cost of deploying four independent models. This demonstrates that ST-CLIP achieves a more balanced trade-off between accuracy and efficiency.

\begin{table}[t]
    \centering
    \footnotesize
    \caption{Comparison of model efficiency across different methods.}
    \begin{tabular}{c|ccc}
        \toprule
        \textbf{Method} & \textbf{FLOPs (G)} & \textbf{\makecell{Inference\\ Latency (ms)}} & \textbf{\makecell{Peak\\ Memory (MiB)}} \\
        \midrule
        CoOp            & 37.9 & 13.4$\pm$1.2 & 506.6 \\
        CoCoOp          & 37.9 & 14.4$\pm$2.2 & 506.6 \\
        CLIP-Adapter    & 37.9 & 18.6$\pm$1.1 & 506.6 \\
        Tip-Adapter-F   & 54.4 & 28.2$\pm$3.1 & 560.2 \\
        ST-CLIP         & 72.1 & 36.7$\pm$4.6 & 683.6 \\
        \bottomrule
    \end{tabular}
    \label{tab:efficiency_comparison}
\end{table}

\section{Parameter Analysis}\label{sec:parameter}

In \name, we select ViT-B/32 as the vision backbone, which is also adopted in the baselines~\cite{zhou2022conditional, zhang2021tip}. 
Additionally, we set the length of learnable prompts to 16 and place the class-specific word $[\mathrm{CLASS}]$ at the end of the prompt, which follows the settings in the baselines~\cite{zhou2022learning, zhou2022conditional}. 
Further analysis of these parameters is provided below.

\vspace{-0.3cm}
\subsection{Length of Learnable Context Vectors}

\begin{figure}[tbp]
    \centering
    \includegraphics[width=1\linewidth]{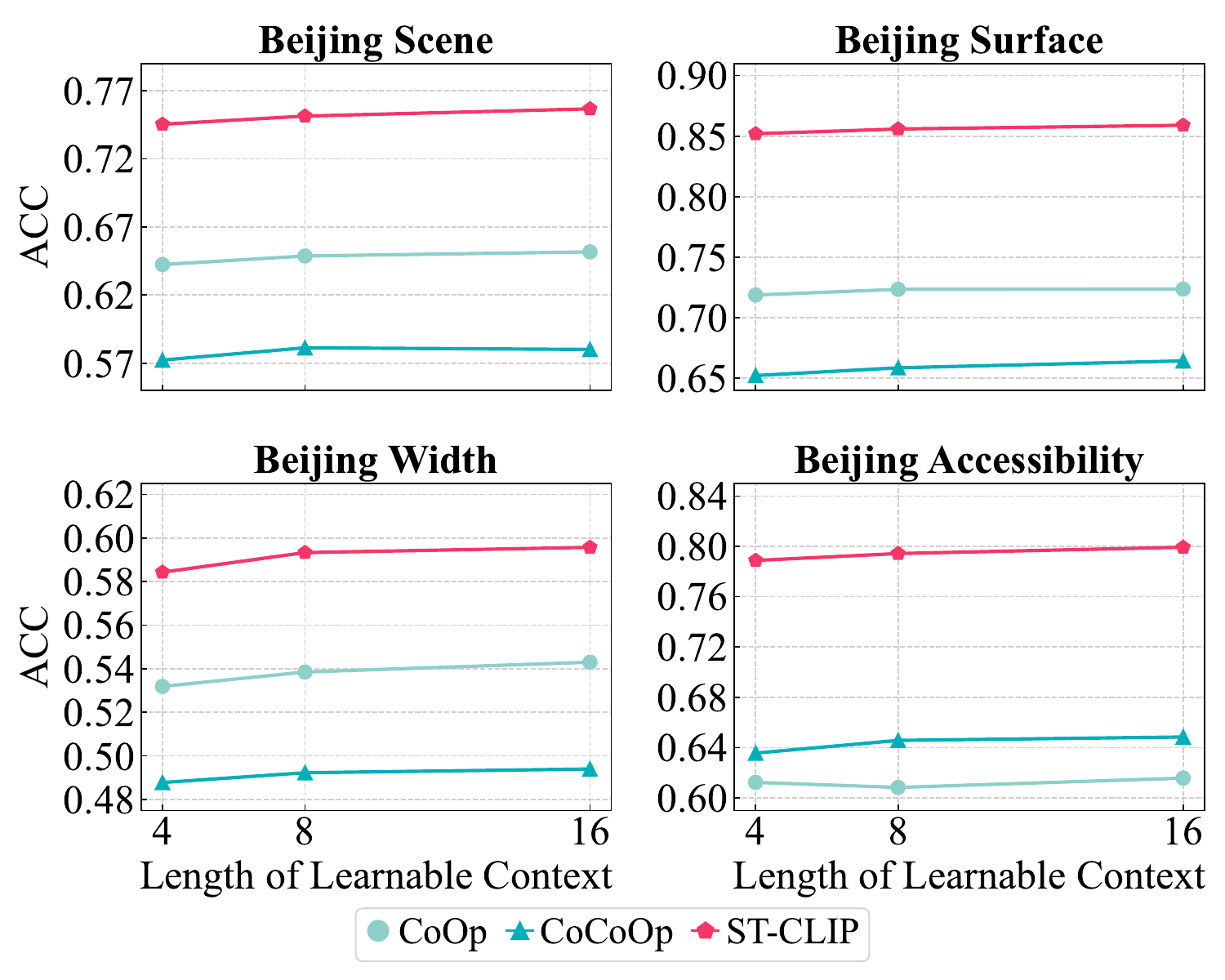}
    \caption{Influence of the length of learnable context vectors on Beijing dataset for four tasks.}
    \label{context}
\end{figure}

To explore the impact of the length of learnable context vectors on the performance, we conduct experiments on the \emph{Beijing} dataset. Since CoOp and CoCoOp are also affected by the length of learnable context vectors, we compare their performance with ST-CLIP under the conditions of 4, 8, and 16 vectors respectively. 
Fig.~\ref{context} illustrates the experimental results obtained from the \emph{Beijing} dataset. It is evident that as the length of learnable context vectors increases, the performance of the three models in different classification tasks shows slight improvement. 
CoOp outperforms CoCoOp in the scene, surface, and width datasets, but it performs worse than CoCoOp in the accessibility dataset. Notably, ST-CLIP consistently achieves the best performance compared to both CoOp and CoCoOp across all datasets. When the length of learnable context vectors reaches 16, ST-CLIP demonstrates the highest performance among the models.

\vspace{-0.3cm}
\subsection{Position of Class-specific Word [CLASS]}

\begin{figure}[htbp]
    \centering
    \includegraphics[width=1\linewidth]{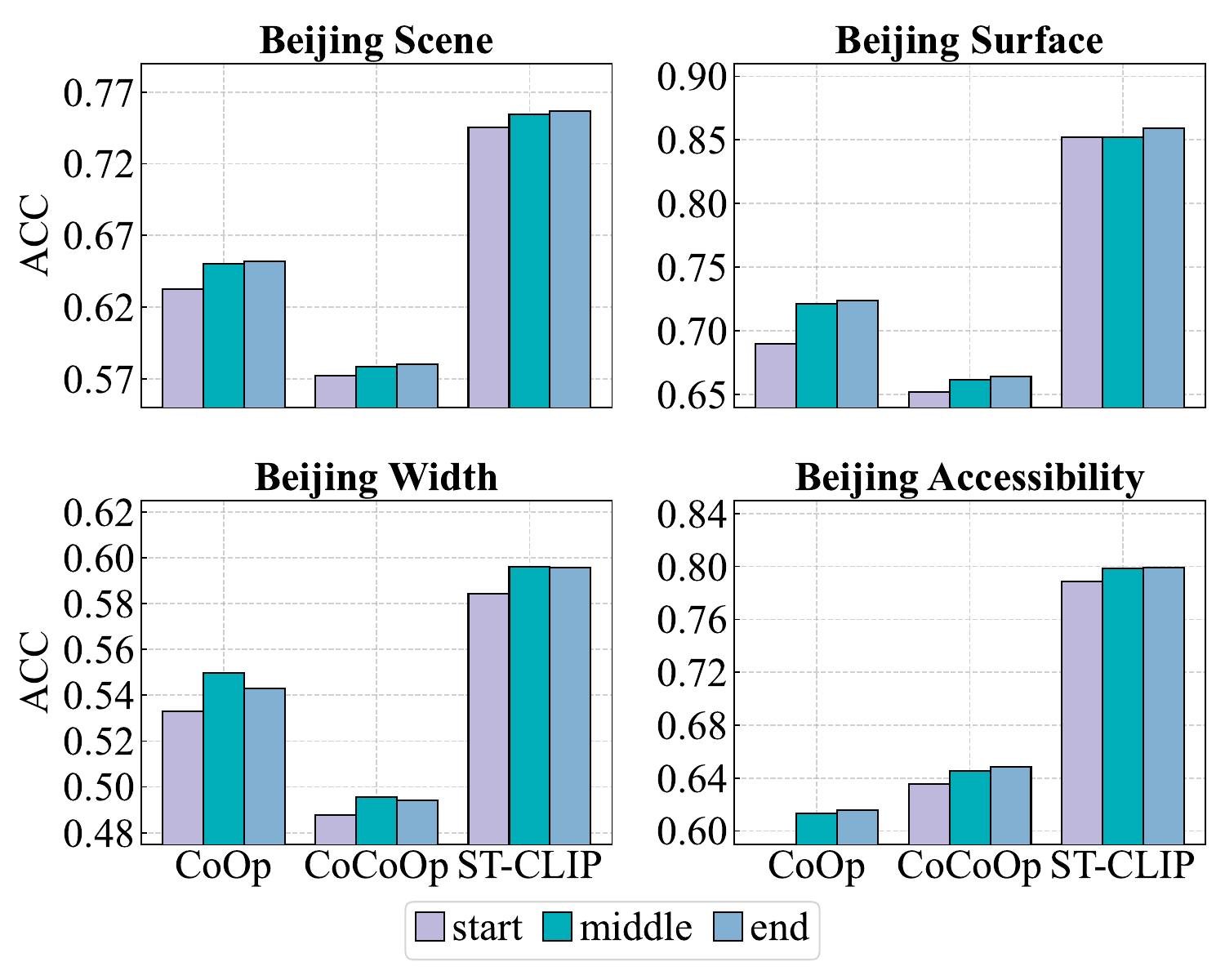}
    \caption{Influence of the position of class-specific word [CLASS] on Beijing dataset for four tasks.}
    \label{position}
\end{figure}

Similarly, we compare CoOp and CoCoOp with ST-CLIP to investigate the impact of the position of the class-specific word [CLASS] on model performance. In our experiments, we test three positions: \emph{start}, \emph{middle}, and \emph{end}. Specifically, the \emph{start} position places [CLASS] at the beginning of the prompt; the \emph{middle} position places [CLASS] between two sets of 8 learnable vectors; and the \emph{end} position places all 16 learnable vectors before [CLASS].  
As shown in Fig.~\ref{position}, the \emph{start} position consistently performs the worst, likely because this design does not align with common linguistic patterns, making it harder for the learnable context to fully account for the preceding class word. The \emph{middle} and \emph{end} positions yield comparable performance, with the \emph{middle} slightly better for the road width classification task, while the \emph{end} performs better for the remaining tasks. Notably, ST-CLIP achieves the best results across all experimental settings, confirming its robustness to prompt position variations.

\begin{table*}[htbp]
    \centering
    \small
    \caption{Qualitative case studies of attention visualization with interpretability analysis.}
    \begin{tabular}{>{\centering\arraybackslash}m{0.15\textwidth} 
                    >{\centering\arraybackslash}m{0.15\textwidth} 
                    >{\centering\arraybackslash}m{0.15\textwidth} |
                    >{\centering\arraybackslash}m{0.22\textwidth} |
                    >{\centering\arraybackslash}m{0.20\textwidth}}
        \toprule
        \textbf{Image} & \textbf{Background Scene} & \textbf{Foreground Road} & \textbf{Description} & \textbf{Error Cause} \\
        \midrule
        \includegraphics[width=0.14\textwidth]{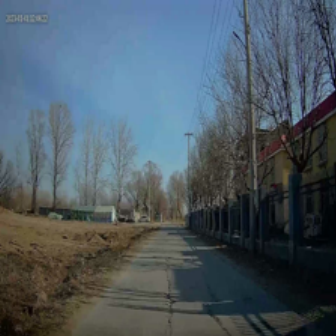} &
        \includegraphics[width=0.15\textwidth]{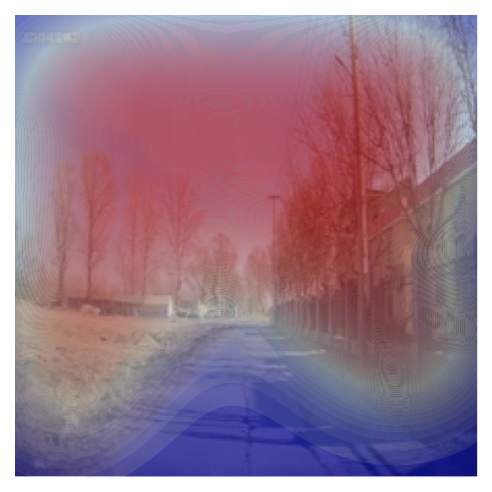} &
        \includegraphics[width=0.15\textwidth]{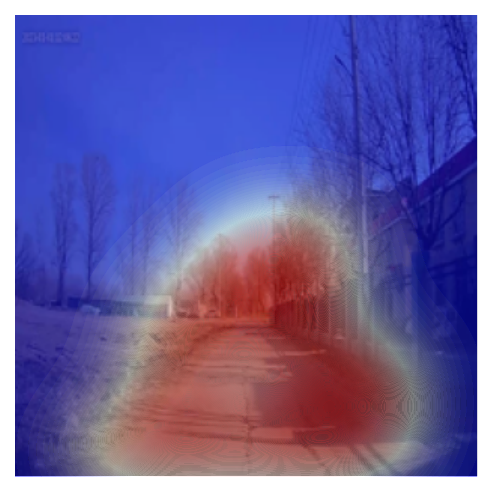} &
        \noindent\justifying The road is in the {\color{green}{\textit{field}}} scene. The surface is {\color{green}{\textit{broken}}}. The width is {\color{green}{\textit{narrow}}}. It’s {\color{green}{\textit{easy}}} to pass through. & 
        \multirow{2}{*}[-4em]{/} \\
        \cmidrule(lr){1-4}
        \includegraphics[width=0.14\textwidth]{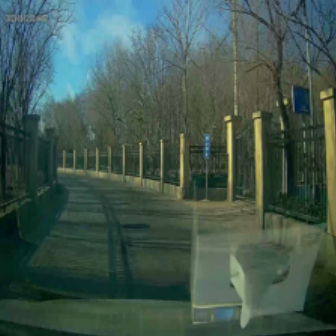} &
        \includegraphics[width=0.15\textwidth]{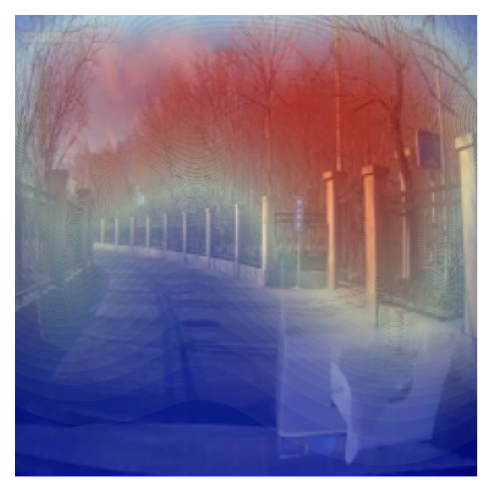} &
        \includegraphics[width=0.15\textwidth]{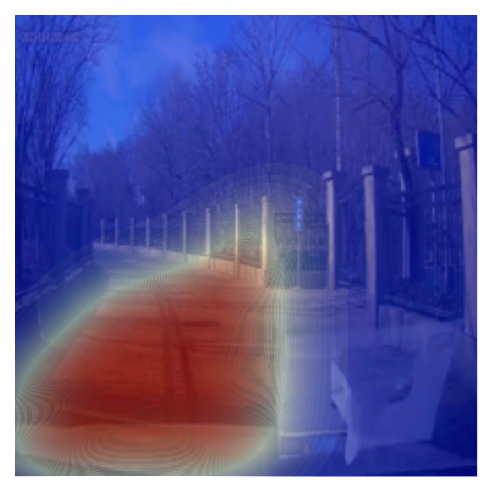} &
        \noindent\justifying The road is in the {\color{green}{\textit{field}}} scene. The surface is {\color{green}{\textit{normal}}}. The width is {\color{green}{\textit{normal}}}. It’s {\color{green}{\textit{easy}}} to pass through. & \\
        \midrule
        \includegraphics[width=0.14\textwidth]{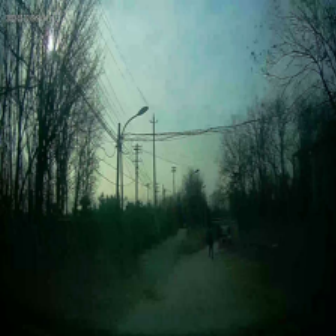} &
        \includegraphics[width=0.15\textwidth]{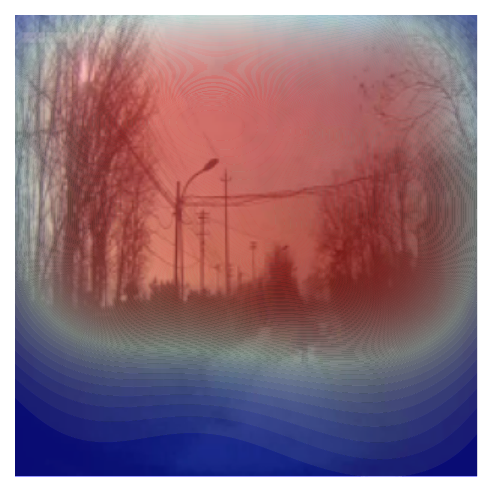} &
        \includegraphics[width=0.15\textwidth]{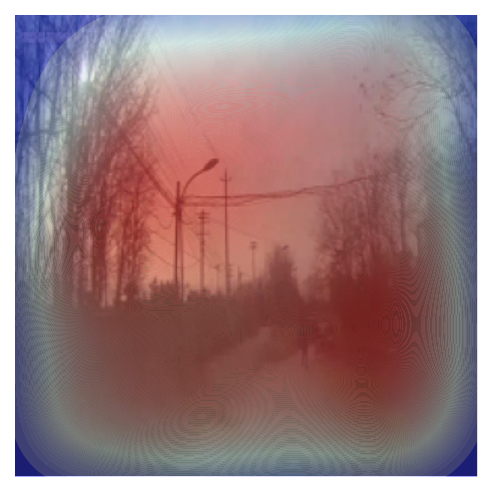} &
        \noindent\justifying The road is in the {\color{green}{\textit{field}}} scene. The surface is {\color{red}{\textit{normal}}}. The width is {\color{red}{\textit{normal}}}. It’s {\color{red}{\textit{easy}}} to pass through. & 
        \multirow{2}{*}[-3em]{\parbox{0.20\textwidth}{\raggedright Lighting condition: glare and low visibility caused the model to misinterpret the foreground road area.}} \\
        \cmidrule(lr){1-4}
        \includegraphics[width=0.14\textwidth]{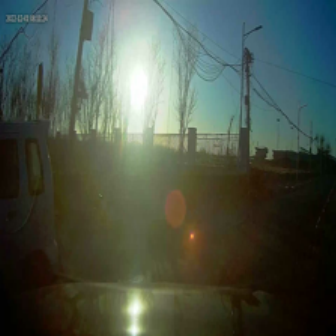} &
        \includegraphics[width=0.15\textwidth]{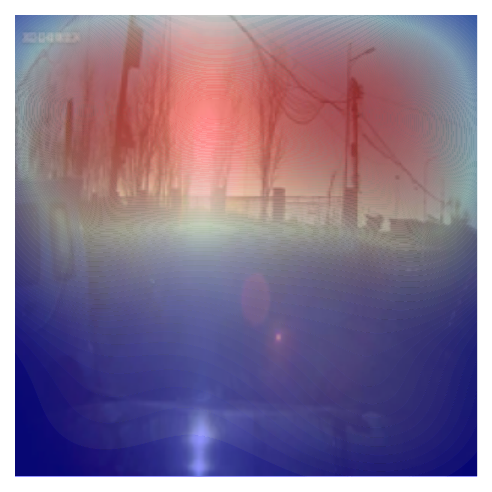} &
        \includegraphics[width=0.15\textwidth]{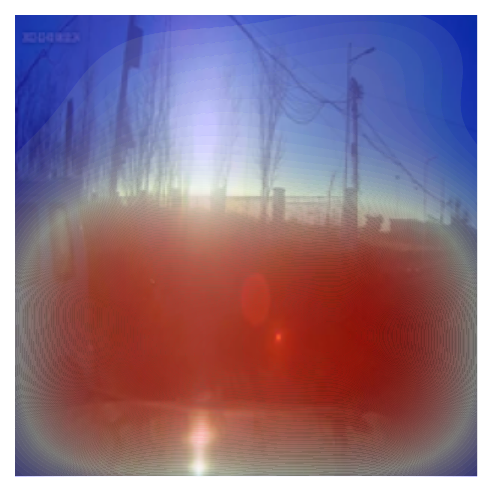} &
        \noindent\justifying The road is in the {\color{green}{\textit{field}}} scene. The surface is {\color{green}{\textit{soil}}}. The width is {\color{red}{\textit{normal}}}. It’s {\color{red}{\textit{easy}}} to pass through. & \\
        \midrule
        \includegraphics[width=0.14\textwidth]{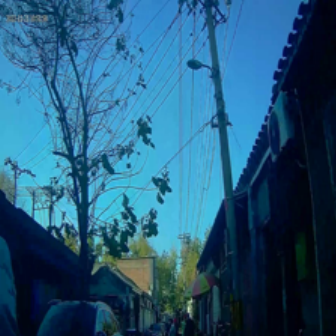} &
        \includegraphics[width=0.15\textwidth]{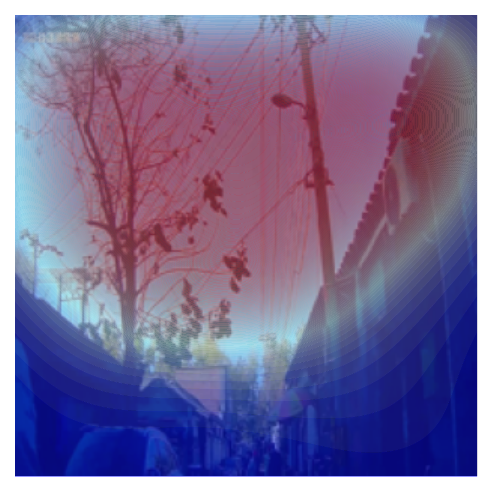} &
        \includegraphics[width=0.15\textwidth]{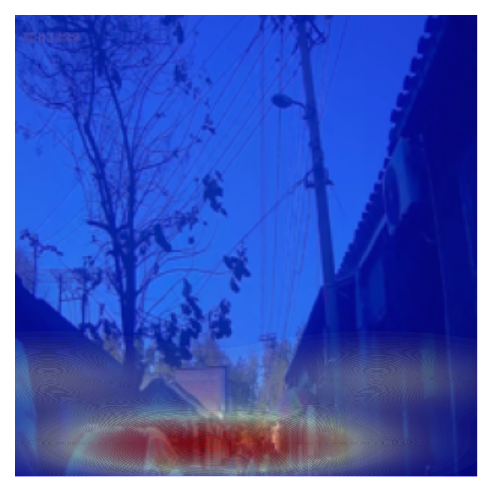} &
        \noindent\justifying The road is in the {\color{green}{\textit{field}}} scene. The surface is {\color{red}{\textit{unknown}}}. The width is {\color{green}{\textit{narrow}}}. It’s {\color{green}{\textit{hard}}} to pass through. & 
        \multirow{2}{*}[-2em]{\parbox{0.20\textwidth}{\raggedright Viewpoint limitation: road slope or camera angle prevented effective capture of the road surface, leading to uncertainty in surface estimation.}} \\
        \cmidrule(lr){1-4}
        \includegraphics[width=0.14\textwidth]{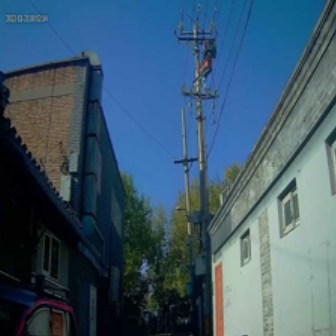} &
        \includegraphics[width=0.15\textwidth]{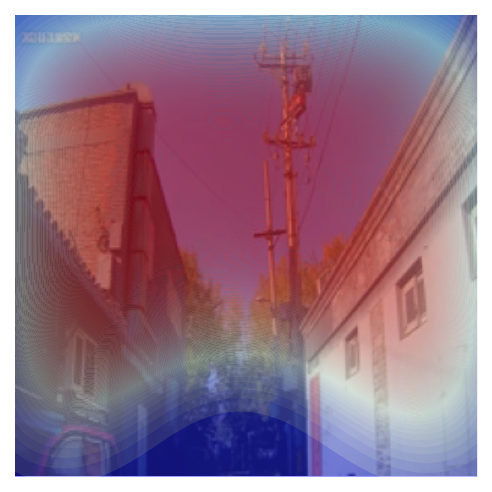} &
        \includegraphics[width=0.15\textwidth]{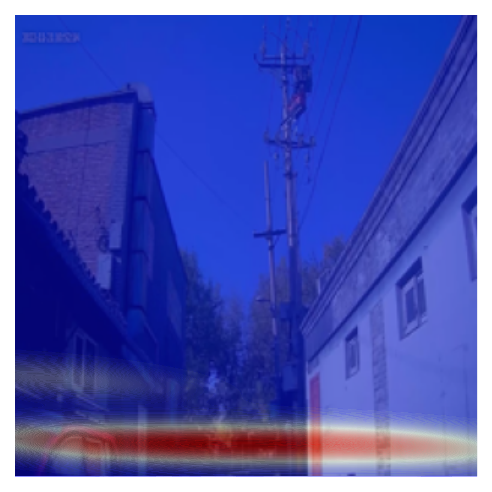} &
        \noindent\justifying The road is in the {\color{green}{\textit{field}}} scene. The surface is {\color{green}{\textit{unknown}}}. The width is {\color{red}{\textit{normal}}}. It’s {\color{green}{\textit{hard}}} to pass through. & \\
        \midrule
        \includegraphics[width=0.14\textwidth]{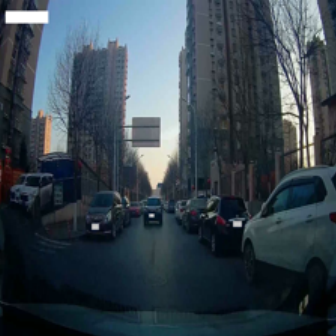} &
        \includegraphics[width=0.15\textwidth]{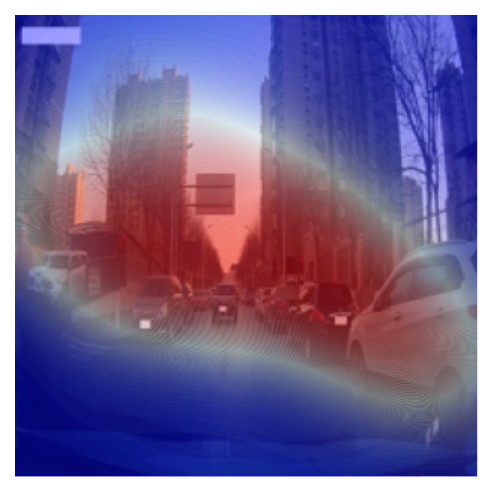} &
        \includegraphics[width=0.15\textwidth]{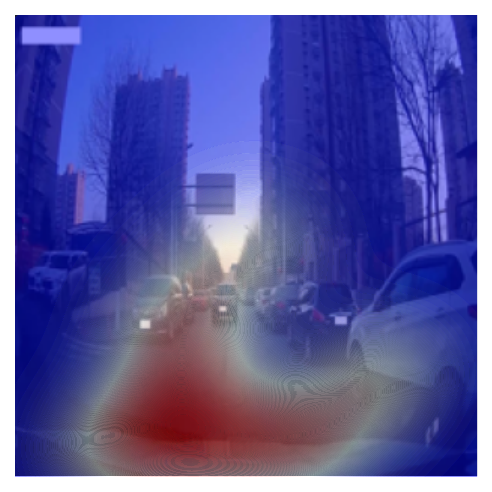} &
        \noindent\justifying The road is in the {\color{green}{\textit{vehicles}}} scene. The surface is {\color{green}{\textit{normal}}}. The width is {\color{red}{\textit{narrow}}}. It’s {\color{green}{\textit{easy}}} to pass through. & 
        \multirow{2}{*}[-3em]{\parbox{0.20\textwidth}{\raggedright Scene ambiguity: the heavy presence of vehicles in the scene caused the model to misinterpret both the scene type and the road width.}} \\
        \cmidrule(lr){1-4}
        \includegraphics[width=0.14\textwidth]{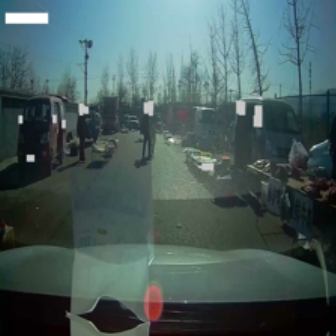} &
        \includegraphics[width=0.15\textwidth]{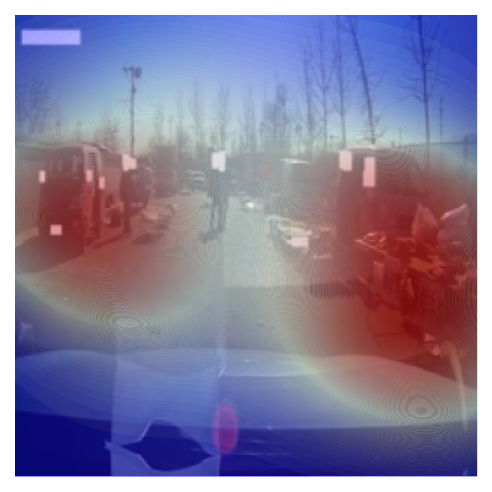} &
        \includegraphics[width=0.15\textwidth]{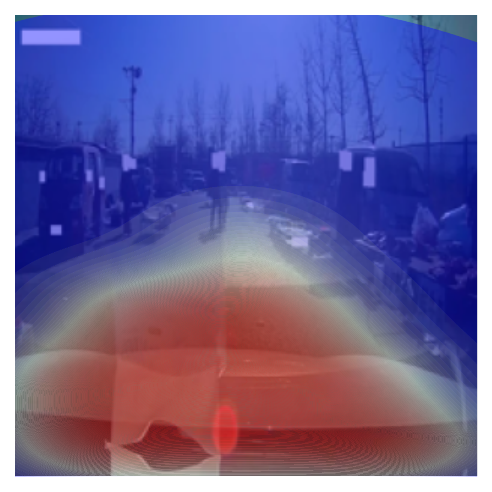} &
        \noindent\justifying The road is in the {\color{red}{\textit{vehicles}}} scene. The surface is {\color{green}{\textit{normal}}}. The width is {\color{green}{\textit{normal}}}. It’s {\color{green}{\textit{easy}}} to pass through. & \\
        \bottomrule
    \end{tabular}
    \label{tab:qualitative_case}
\end{table*}

\vspace{-0.3cm}
\section{FMM Algorithm}
Fast Map Matching (FMM)~\cite{yang2018fast} is an efficient algorithm designed to perform map matching on large-scale road networks. It integrates HMM-based inference with precomputed spatial indexing structures to achieve fast and accurate trajectory-to-road network alignment. It typically follows a three-step process to perform road network matching:

\begin{itemize}
    \item \textit{Candidate Search Using Spatial Indexing:}
    An R-tree or grid-based index efficiently retrieves candidate road segments near each GPS point, considering both Euclidean and network distances.
    
    \item \textit{Transition Probability Computation:}  
    Using an HMM framework, transition probabilities between candidate segments are computed based on shortest-path distances, ensuring realistic vehicle movements.  
    
    \item \textit{Viterbi-Based Path Inference:}  
    The Viterbi algorithm infers the most probable sequence of road segments by balancing GPS observation errors with network constraints. This achieves a globally optimal match while maintaining high efficiency and accuracy, making FMM suitable for large-scale trajectory data.  
\end{itemize}

By leveraging efficient spatial indexing and optimized HMM-based inference, FMM achieves a balance between computational efficiency and matching accuracy, making it suitable for large-scale trajectory data processing.

\begin{table}[t]
    \small
    \centering
    \caption{Experimental results on extended scenes.}
    \begin{tabular}{c|c|c|c}
    \toprule
    Scene Class & \# Total & \# Correct & Accuracy \\
    \midrule
    Expressway & 761 & 724 & 0.951 \\ 
    Construction Road & 797 & 784 & 0.984 \\
    Avenue & 121 & 87 & 0.719 \\
    \bottomrule
    \end{tabular}
    \label{tab:scalability}
\end{table}

\section{Case Study}
To further strengthen the interpretability analysis, we conducted qualitative case studies that visualize and analyze how the model makes decisions in both successful and failure scenarios.
We present a richer set of good cases (where the model makes correct predictions) and bad cases (where it fails). For each case, we include: the raw traffic scene image, the corresponding background scene attention heatmap, and the foreground road attention heatmap. 
These visualizations highlight which image regions contributed most strongly to predictions of different aspects.

As shown in Table~\ref{tab:qualitative_case}, good cases demonstrate that the bi-level attention mechanism consistently focuses on semantically meaningful regions (e.g., the sky and trees for “scene,” or the road surface for “width”). In contrast, failure cases expose situations where attention is distracted or misdirected, thereby revealing the model’s weaknesses. We identified three recurring patterns that often lead to misclassification:

\begin{itemize}
    \item Lighting condition: Strong glare or low illumination reduces visibility of critical road regions, causing the model to focus on irrelevant patches in the foreground and misinterpret the road surface.
    \item Viewpoint limitation: When the road is captured at a steep slope or from an unusual camera angle, road details become indistinct, leading to uncertainty in classification.
    \item Scene ambiguity: In heavily congested traffic scenes, vehicles occlude large parts of the foreground road, introducing confusion that sometimes causes the model to conflate scene type with road width.
\end{itemize}

These case studies provide human-interpretable insights into why the model may fail, going beyond numerical metrics. By combining narrative captions with visual attention heatmaps, the interpretability of the framework is enhanced, and the decision-making process becomes more transparent.

\vspace{-0.3cm}
\section{Scalability Test}
Due to the excellent scalability, ST-CLIP can easily adapt to diverse traffic scenarios, whether by expanding the number of aspects or increasing the number of class-specific words for a particular aspect.
We extend three additional class-specific words for the ``Scene'' aspect:
\begin{itemize}
    \item \textit{Expressway}: It represents road segments with multiple lanes and a high average speed.
    \item \textit{Construction Road}: It represents road segments under construction, typically featuring barriers or traffic cones.
    \item \textit{Avenue}: It represents road segments shaded by trees.
\end{itemize}
We integrate these data and conduct few-shot experiments using the same settings and report the results for these three scenes in Table~\ref{tab:scalability}. The experimental results indicate that our model can generalize to diverse traffic scenes, particularly those with distinct features such as expressways and construction roads. The recognition accuracy exceeds 95\% in these scenarios. For avenues, there are fewer images available, and due to factors such as lighting, the image clarity is lower, resulting in a lower classification accuracy.

\bibliographystyle{IEEEtran}
\bibliography{reference}

\begin{IEEEbiography}[{\includegraphics[width=1in,height=1.25in,clip,keepaspectratio]{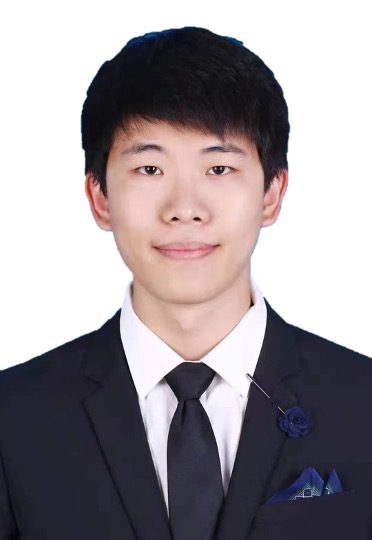}}]{Jingtian Ma}
received the BEng degree in computer science from the Beihang University, China, in 2019. He is currently working toward the PhD degree in the School of Computer Science and Engineering, Beihang University. His research primarily focuses on spatio-temporal data mining, multimodal learning and large language models.
\end{IEEEbiography}

\begin{IEEEbiography}[{\includegraphics[width=1in,height=1.25in,clip,keepaspectratio]{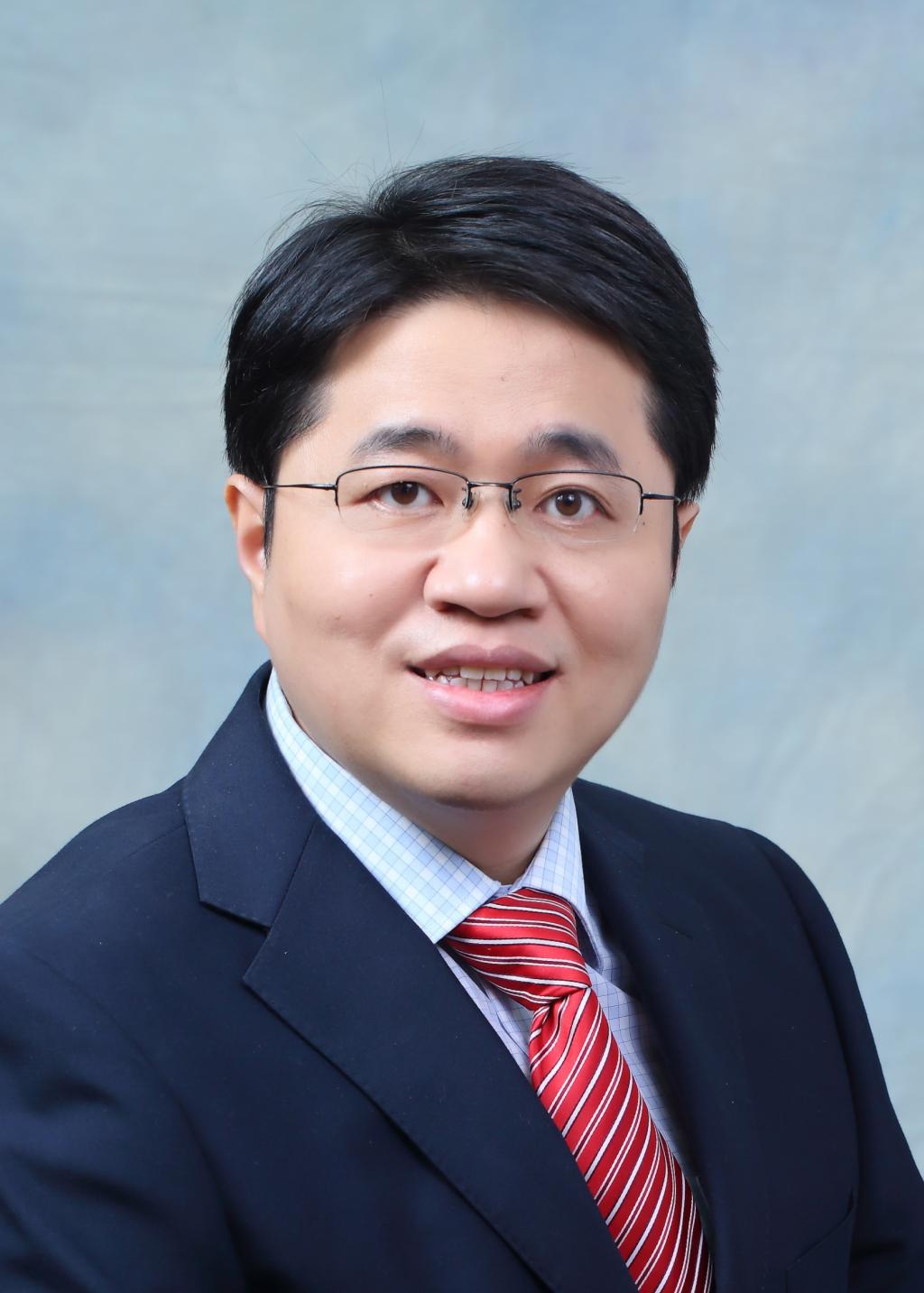}}]{Jingyuan Wang}
received the Ph.D. degree from the Department of Computer Science and Technology, Tsinghua University, Beijing, China. He is currently an Associate Professor of School of Computer Science and Engineering, Beihang University, Beijing, China. His is also the leader of Beihang Interest Group on SmartCity (BIGSCity), and Vice Director of the Beijing City Lab (BCL). He published more than 20 papers on top journals and conferences, as well as named inventor on several granted US patents. His general area of research is data mining and machine learning, with special interests in smart cities.
\end{IEEEbiography}

\begin{IEEEbiography}[{\includegraphics[width=1in,height=1.25in,clip,keepaspectratio]{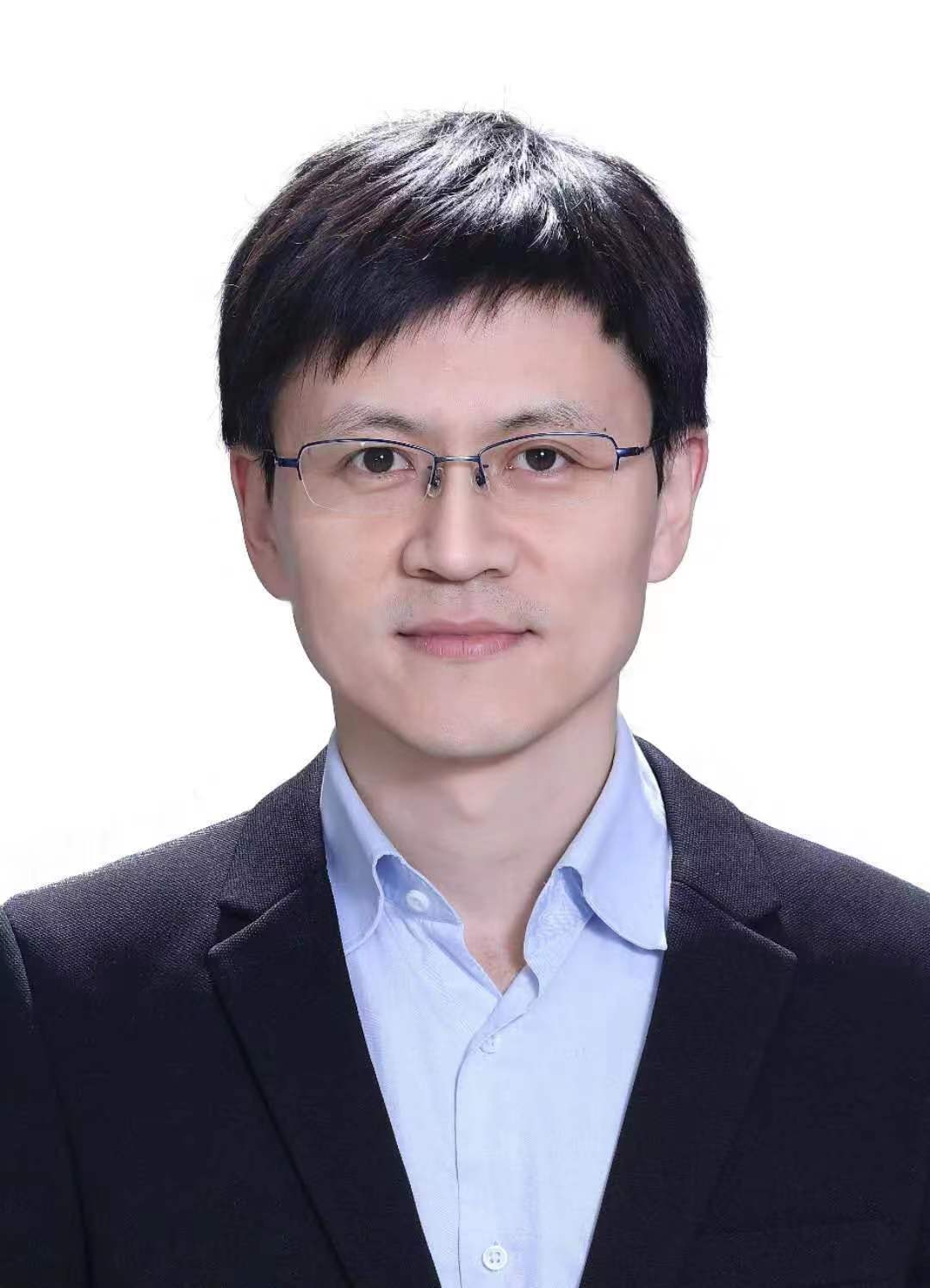}}]{Wayne Xin Zhao}
received the PhD degree from Peking University in 2014. He is currently a tenured associated professor in Gaoling School of Artificial Intelligence, Renmin University of China. His research interests are web text mining and natural language processing. He has published a number of papers in international conferences and journals such as ACL, SIGIR, SIGKDD, WWW, ACM TOIS, and IEEE TKDE. He is a member of the IEEE.
\end{IEEEbiography}

\begin{IEEEbiography}[{\includegraphics[width=1in,height=1.25in,clip,keepaspectratio]{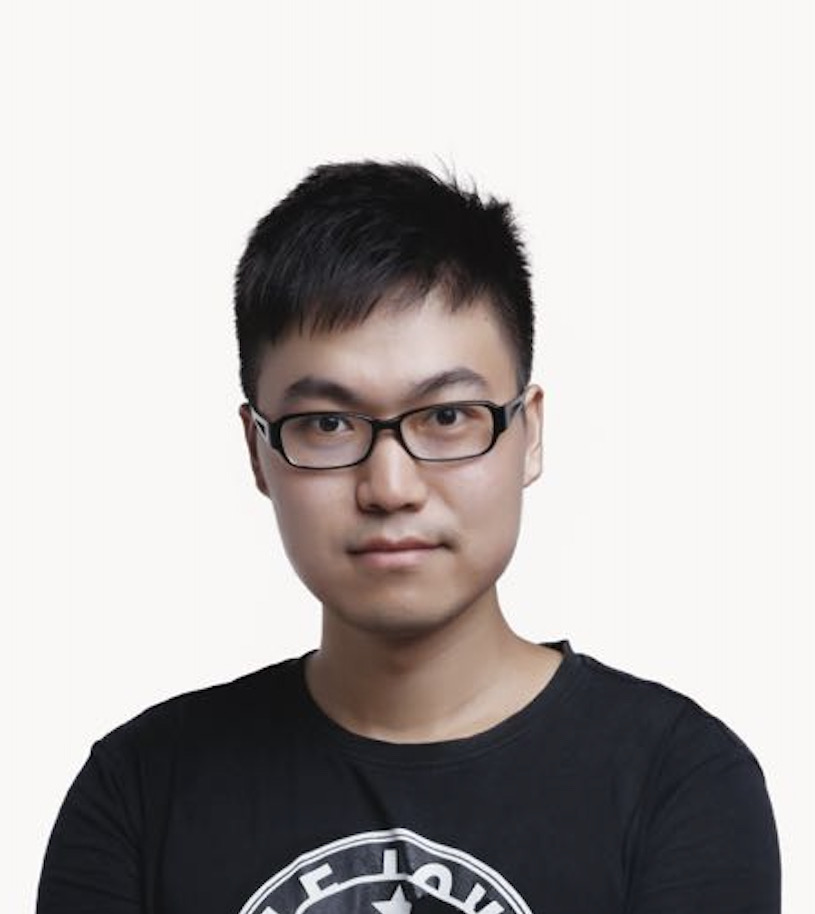}}]{Guoping Liu}
received the M.S. degree from Zhejiang University in 2016. He is currently an Algorithm Engineer with Didi Chuxing. His research interests include spatio-temporal anomaly detection, trajectory pattern mining, user travel portrait, and road network construction.
\end{IEEEbiography}

\begin{IEEEbiography}[{\includegraphics[width=1in,height=1.25in,clip,keepaspectratio]{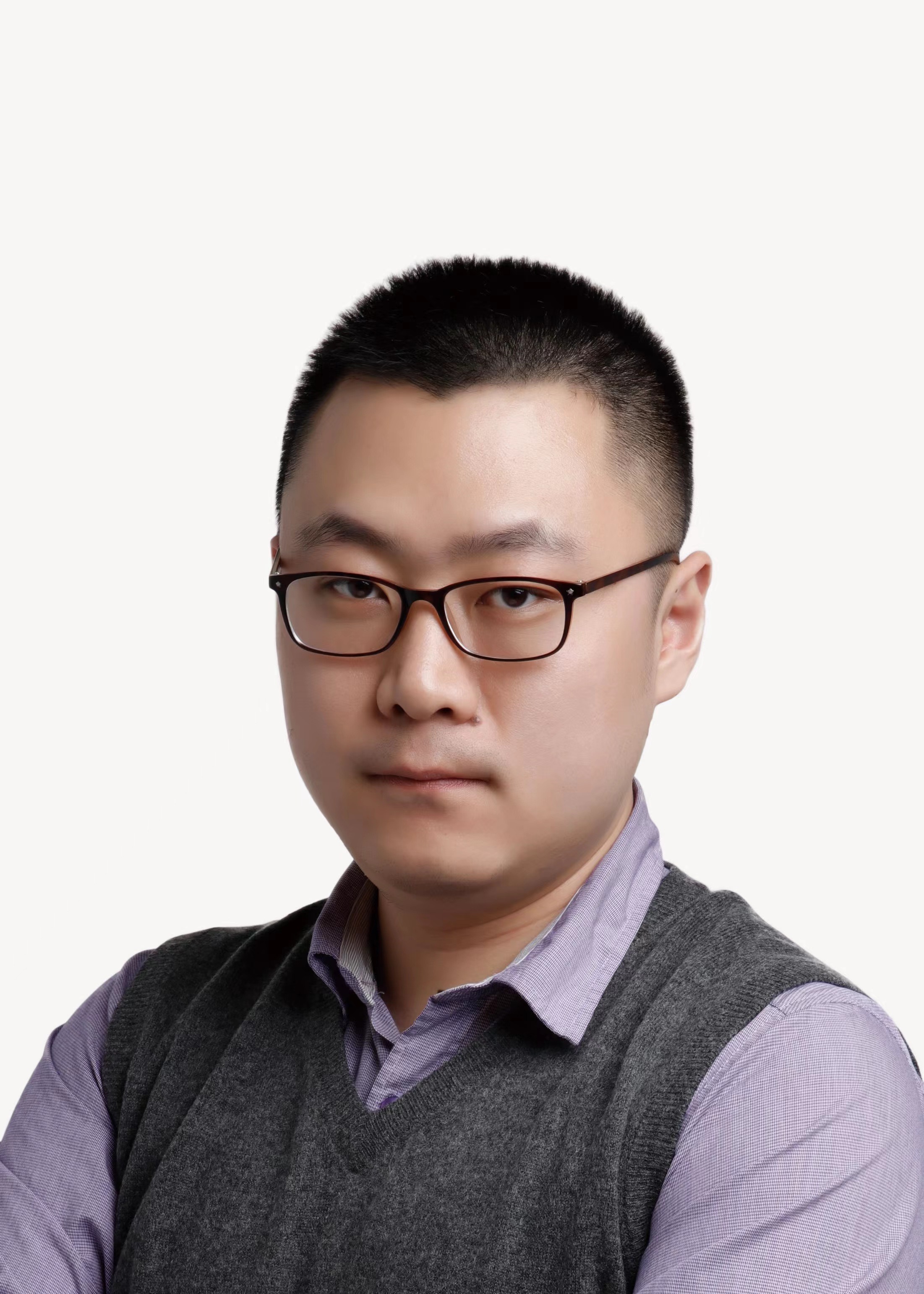}}]{Xiang Wen}
received the M.S. degree in computer science from Peking University in 2012. He is currently an Expert Engineer with the Business Department of Didi Map and Bus, Didi Chuxing. He majors in trajectory mining and map security feature platform.
\end{IEEEbiography}

\end{document}